\documentclass[preprint,12pt]{aastex}

\newcommand{\degree}{\mbox{$^{\circ}$}}

\newcommand{\as}{\mbox{\arcsec}}

\newcommand{\lsun}{\mbox{L$_\odot$}}
\newcommand{\msun}{\mbox{M$_\odot$}}

\input{epsf}

\def\plotfiddle#1#2#3#4#5#6#7{\centering \leavevmode
\vbox to#2{\rule{0pt}{#2}}
\includegraphics{#1}}

\newcommand{\msunmyr}{\mbox{M$_\odot$ Myr$^{-1}$}}

\begin{document}


\title {\bf The Spitzer c2d Survey of Large, Nearby, Insterstellar Clouds. IX. The Serpens YSO Population As Observed With IRAC and MIPS}
\author{Paul Harvey\altaffilmark{1},
Bruno Mer\'{\i}n\altaffilmark{2},
Tracy L. Huard\altaffilmark{3},
Luisa M. Rebull\altaffilmark{4},
Nicholas Chapman\altaffilmark{5},
Neal J. Evans II\altaffilmark{1},
Philip C. Myers\altaffilmark{3}
}

\altaffiltext{1}{Astronomy Department, University of Texas at Austin, 1 University Station C1400, Austin, TX 78712-0259;  pmh@astro.as.utexas.edu, nje@astro.as.utexas.edu}
\altaffiltext{2}{Research and Scientific Support Dept., ESTEC (ESA), Keplerlaan, 1, PO Box 299, 2200 AG Noordwijk, The Netherlands and Leiden Observatory, Leiden University, Postbus 9513, 2300 RA Leiden, The Netherlands; bmerin@rssd.esa.int}
\altaffiltext{3}{Smithsonian Astrophysical Observatory, 60 Garden Street, MS42, Cambridge, MA 02138; thuard@cfa.harvard.edu, pmyers@cfa.harvard.edu}
\altaffiltext{4}{Spitzer Science Center, MC 220-6, Pasadena, CA 91125; rebull@ipac.caltech.edu}
\altaffiltext{5}{Astronomy Department, University of Maryland, College Park, MD 20742; chapman@astro.umd.edu}

\begin{abstract}

We discuss the results from the combined IRAC and MIPS c2d Spitzer Legacy observations of the Serpens 
star-forming region.  In particular we present a set of criteria for isolating bona fide young stellar
objects, YSO's, from the extensive background contamination by extra-galactic objects.  We then discuss
the properties of the resulting high confidence set of YSO's.  We find 235 such objects in the 0.85 deg$^2$
field that was covered with both IRAC and MIPS.  An additional set of 51 lower confidence YSO's outside
this area is identified from the MIPS data combined with 2MASS photometry.  To understand the
properties of the circumstellar material that produces the observed infrared emission, we describe two
sets of results, the use of color-color diagrams to compare our observed source properties with those
of theoretical models for star/disk/envelope systems and our own modeling of the subset of our objects that appear
to be well represented by a stellar photosphere plus circumstellar disk.  These objects exhibit a very
wide range of disk properties, from many that can be fit with actively accreting disks to some with both
passive disks and even possibly debris disks.
We find that the luminosity function of YSO's in Serpens extends down to at least a few $\times 10^{-3}$ \lsun\
or lower for an assumed distance of 260 pc. The lower limit may be set by our inability to distinguish YSO's from extra-galactic
sources more than by the lack of YSO's at very low luminosities.  We find no evidence for variability
in the shorter IRAC bands between the two epochs of our data set, $\Delta t \sim$ 6 hours.  A spatial
clustering analysis shows that the nominally less-evolved YSO's are more highly clustered than the
later stages and that the background extra-galactic population can be fit by the same two-point
correlation function as seen in other extra-galactic studies.  We also present a table of matches
between several previous infrared and X-ray studies of the Serpens YSO population and our
Spitzer data set.
The clusters in Serpens have a very high surface density of YSOs, primarily
with SEDs suggesting extreme youth. The total number of YSOs, mostly Class II,
is greater in the region outside the clusters.

\end{abstract}


\section{Introduction}\label{intro}

The Serpens star-forming cloud is one of five such large clouds selected for observation as part of
The Spitzer Legacy project ``From Molecular Cores
to Planet-forming Disks'' (c2d) \citep{evans03}.  Previous papers in this series have
described the observational results in the Serpens Cloud as seen with IRAC \citep{harv06}(Paper I) and
MIPS \citep{harv07} as well as some of the other clouds \citep{jorg06,rebull06}.  
In this paper we examine how the combination of the IRAC and
MIPS data together with other published results on this region can be used to find and characterize
a highly reliable catalog of young stellar objects (YSO's) in the surveyed area. 
With the combination of broad wavelength coverage and amazing depth of Spitzer's sensitivity
we are able to probe to both extremely low luminosity limits for YSO's and to cover a very
wide range in dust emission, both in optical depth and in range of emitting temperatures.
The Spitzer wavelength region is particularly well tuned for sensitivity to dust at temperatures
appropriate for solar-system size disks around young stars.

The region of the Serpens Cloud mapped in our survey is an area rich in star formation. \citet{eiroa07} have
extensively reviewed studies at a variety of wavelengths of this area.  There is evidence
from previous observations of strong clustering \citep{testi00,ts98}, dense sub-mm cores \citep{cas93,enoch07}, 
and high-velocity
outflows \citep{ze99,dav99}.  Pre-Spitzer infrared surveys of the cloud have been made by IRAS \citep{zhang88a,zhang88b} and
ISO \citep{kaas04,djup06} as well as the pioneering ground-based surveys that first identified it as an
important region of star formation \citep{svs76}.
Using MIPS and all four bands of IRAC the c2d program has mapped
a 0.85 deg$^2$ portion of this cloud that includes a very well-studied
cluster of infrared and sub-millimeter sources \citep{ec92,hds99,hb96,harv84,ts98}.  
At its distance of $260\pm10$ pc \citep{stra96} this corresponds to an area of about
2.5 $\times$ 9 pc.  
In paper I we identified at least two main
centers of star formation as seen by Spitzer in this cloud, that we referred to as
Cluster A and B.  Cluster A is the very well-studied grouping also commonly referred to as the
Serpens Core.  Cluster B was the subject of a recent multi-wavelength study by
\citet{djup06}, who referred to it as the Serpens G3-G6 cluster.

The 235 YSO's with high signal-to-noise that we have catalogued constitute a sufficiently large number
that we can examine statistically the numbers of objects
in various evolutionary states and the range of disk properties for different classes of YSO's.
We characterize the circumstellar material with color-color diagrams that
allow comparison with other recent studies of star-forming regions, and we model the
energy distributions of the large number of YSO's that appear to be star$+$circumstellar disk
systems.  We are able to construct the YSO luminosity function for Serpens since we have complete
spectral coverage for all the sources over the range of wavelengths where their luminosity
is emitted, and we are able to characterize the selection effects inherent in the luminosity
function from comparison with the publicly available and signicantly deeper SWIRE survey \citep{swire}.
We find that the population of YSO's extends down to luminosities below $10^{-2}$ \lsun, and we
discuss the significance of this population.  Our complete coverage in wavelength and luminosity
space also permits us to discuss the spatial distribution of YSO's to an unprecedented completeness
level.  We note also that, unlike the situation discussed by \citet{jorg06} for Perseus, in Serpens
there do not appear to be any very deeply embedded YSO's that are not found by our YSO selection criteria.

In \S\ref{obs} we briefly review the observational details of this program and then in
\S{\ref{selection}} we describe in detail the process by which we identify YSO's
and eliminate background contaminants.  
In \S\ref{vary} we describe a search for variability in our dataset.
We compare our general results
on YSO's with those of earlier studies of the Serpens star-forming region in \S\ref{compare}.
We discuss in
\S\ref{lowlum} the YSO luminosity function in Serpens and, in particular, the low
end of this function.  
We next analyze the spatial distribution
of star formation in the surveyed area  and compare it to the distribution of dust extinction as
derived from our observations in \S\ref{cluster}.
In \S\ref{disks} we construct several color-color diagrams that characterize the
global properties of the circumstellar material and show modeling results for
a large fraction of our YSO's that appear to be star$+$disk systems.  
Finally, \S\ref{select} discusses several specific
groups of objects including the coldests YSO's and a previously identified ``disappearing''
YSO. We also mention several
obvious high-velocity outflows from YSO's that will certainly be the subject of
further study.

\section{Observations}\label{obs}

The parameters of our observations have already been described in
detail in Paper I for IRAC and in a companion study \citep{harv07} for MIPS.  We summarize here some
of the issues most relevant to this study.

\begin{figure}
\plotfiddle{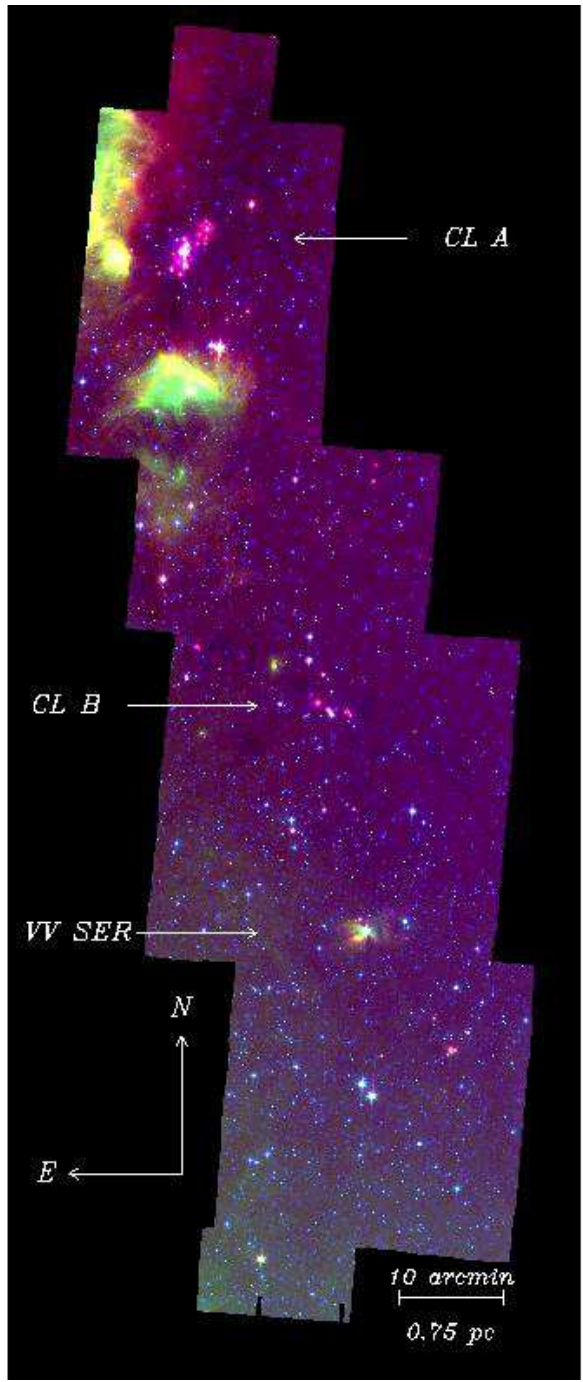}{7.0in}{0}{70}{70}{-200}{-10}
\figcaption{\label{3color-fig}
Three color image of the area mapped in Serpens.  The color mapping is: blue/4.5\micron, green/8.0\micron, and
red/24\micron. The locations of Cluster A (Core Cluster), Cluster B (Serpens G3-G6 Cluster), and VV Ser are indicated.}
\end{figure}

We remind the reader that the angular resolution of the Spitzer imaging instruments varies widely
with wavelength, since Spitzer is diffraction limited longward of $\lambda \sim 10$\micron.  In the shorter
IRAC bands the spatial resolution is of order 2\as, while from 24--160\micron, it goes from roughly
5\as\ to 50\as.
The area chosen for mapping was defined by the $A_V > 6$ contour in the 
extinction map of \citet{camb99} and by practical time constraints \citep{evans03}.  
With the exception of a small area of 0.04 deg$^2$ on the northeast edge of the IRAC map, all of the
area mapped with IRAC was also covered with MIPS at both 24 and 70\micron, and most of it
at 160\micron.
As described by \citet{harv07}, some substantial additional area was observed with
MIPS without matching IRAC observations.  In this current paper we restrict our
attention to only the area that was observed from 3.6 to 70\micron, 0.85 deg$^2$.  This entire region
was observed from 3.6 to 24\micron\ at two epochs, with a time separation of several hours
to several days, but at only one epoch at 70\micron.  We also specifically do not include the 160\micron\ observations in our discussion
because they did not cover the entire area and because most of the YSO's in Serpens are too closely
clustered to be distinguishable in the large beam of the 160\micron\ data.  \citet{harv07} discuss briefly
the extended 160\micron\ emission in this region and the four point-like sources found at 160\micron.
Figure \ref{3color-fig} shows the entire area mapped with all four IRAC
bands and MIPS at 24 and 70\micron\ and also indicates the locations of several areas mentioned in the text.

In addition to this area of the Serpens cloud defined by relatively high A$_V$, we also
observed small off-cloud regions around the molecular cloud with relatively low
A$_V$ in order to determine the background star and galaxy counts.  The area of combined IRAC/MIPS
coverage of these off-cloud regions, however, was relatively small and so these observations are not discussed further in this paper.
As detailed below, we have used the much larger and deeper SWIRE \citep{swire} survey to understand
the characteristics of the most serious background contaminants in our maps, the extra-galactic objects.

\section{YSO Selection}\label{selection}

Paper I described a process for classifying infrared objects into several categories: those whose energy
distributions could be well-fitted as reddened stellar photospheres, those that
had a high likelihood of being background galaxies, and those that were viable
YSO candidates.  We have refined this process by combining MIPS and IRAC data
together as well as by producing an improved comparison catalog from the
SWIRE \citep{swire} survey, trimmed as accurately as possible to the c2d sensitivity limits.
As shown by the number counts versus Wainscoat models of the Galactic background toward
Serpens in paper I, nearly all the sources observed in our survey are likely to be background stars.
Because of the high sensitivity of IRAC channels 1 and 2 relative to both 2MASS and to IRAC channels
3 and 4, most of the more than 200,000 sources extracted from our Serpens dataset do not
have enough spectral coverage for any reliable classification algorithm.  In particular, ``only''
34,000 sources had enough spectral information from a combination of 2MASS and IRAC data to
permit a test for consistency with a stellar photosphere-plus-extinction model, and nearly 32,000
of these objects were classified as reddened stellar photospheres.  These normal stellar objects
are not considered further in our discussion and are not plotted on the various color and magnitude
diagrams.  The details of this classification process and the criteria for fitting are described in
detail by \citet{evans07}.

In order to pursue the classification process for YSO's beyond that described in Paper I, we added one
more step to the data processing described there.  This final step was ``band-filling'' the
catalog to obtain upper limits or low S/N detections of objects that were not found in the original
source extraction processing.  This step is described in detail in the delivery documentation for
the final c2d data delivery \citep{evans07}.  In short, though, it involved fixing the position
of the source during an extraction at the position of an existing catalog source and fitting the
image data at that fixed position for two parameters, a background level and source flux, assuming it
was a point source.  For this processing step, the fluxes of all
the originally extracted sources were subtracted from the image first.  In the case of the data discussed in this
paper, the most important contribution of this band-filling step is to give us flux estimates for the YSO
candidates and extra-galactic candidates at 24\micron.  Because our knowledge of the true PSF is imperfect
and because the 24\micron\ PSF is so much larger than the IRAC ones, it was also necessary to examine
these results carefully to be sure a band-filled 24\micron\ flux was not simply the poorly subtracted
wings of a nearby bright source that, in fact, completely masked the source being band-filled.
For the purposes of this paper, all upper limits are given as 5$\sigma$ values.

Three of the sources eventually selected as YSO's by the process described below had 24\micron\
fluxes that were obviously saturated.  These are the objects in Table \ref{yso-table} numbered
127, 137, and 182.  For these three objects we derived the 24\micron\ flux from a fit to the wings
of the source profiles, rather than a fit to the whole profile as is done by the standard c2d
source extraction.

\def\bm{\bigskip}
\def\micron{$\mu$m}
\def\ptsec{$''\mskip-7.6mu.\,$}
\def\ptmin{$'\mskip-5.0mu.\,$}
\def\pth{$^{h}\mskip-7.6mu.\,$}
\def\ptm{$^{m}\mskip-7.6mu.\,$}
\def\pts{$^{s}\mskip-7.6mu.\,$}
\def\etal{et al.}
\def\ltsim{\raisebox{-.4ex}{$\stackrel{<}{\sim}$}}
\def\gtsim{\raisebox{-.4ex}{$\stackrel{>}{\sim}$}}
\def\arcsec{$^{\prime\prime}$}
\def\arcmin{$^{\prime}$}
\def\twocolheads#1{\multicolumn{2}{c}{#1}}
\def\fivecolhead#1{\multicolumn{5}{c}{#1}}
\def\fullcolhead#1{\multicolumn{6}{c}{#1}}
\def\et{\bm\end{table}}
\def\bt{\begin{table}[htbp]\bm}
\def\efig{\end{figure}}
\def\bfig{\begin{figure}[htb]\bm}

\newcommand{\vdag}{(v)^\dagger}

\subsection{Constructing a Control Catalog from Deep Extragalactic SWIRE Observations}\label{resampSWIRE}

We use the IRAC and MIPS images of the ELAIS~N1 field obtained by the SWIRE team
as a control field for understanding the extragalactic population with colors that mimic
those of YSOs present in our Serpens field.  This SWIRE field, expected to contain no
extinction from a molecular cloud and no YSOs, has coverage by both
IRAC and MIPS of 5.31 deg$^2$ and has a limiting flux roughly a factor of four below that of our observations of
Serpens.  The analysis described below is designed to produce a resampled version of the
SWIRE field as it would have been observed with the typical c2d sensitivity and as if it were
located behind a molecular cloud with the range of extinctions observed in Serpens.
The process of simulating these effects is discussed in detail in the
final c2d data delivery documentation \citep{evans07}, but the steps are summarized here.

To avoid effects that may result from differences in data processing, the BCD images for this
SWIRE field were processed by our pipeline in exactly the same way as our own
observations.  Once a bandmerged catalog of SWIRE sources was
constructed, we first simulated the reddening of sources that would occur if Serpens
had been in the foreground of this field.  This reddening was accomplished by 
randomly applying extinction to each SWIRE source according to the extinction 
profile of Serpens, shown in Figure~\ref{histAV}.  For example, $\sim$23\% of SWIRE 
sources were randomly selected and visual extinctions in the range 6.5 $\leq\ A_V <$ 7.5
were applied, $\sim$19\% of sources were extincted by extinctions in the range 7.5 $\leq\ A_V <$ 8.5,
and so forth.  The extinctions were applied to each of the infrared bands according to the 
extinction law appropriate for molecular clouds and cores (Huard et al., in prep.).

Second, we degraded the sensitivity of the reddened SWIRE photometry to match that of our
Serpens observations.  This was accomplished by matching the detection rates
as a function of magnitude in each of the bands.  The 90\% completeness limits of
the Serpens observations are approximately 16.6, 15.6, 15.0, 16.6, 16.2, 15.2, 13.4,
and 9.6 mag at J, H, K, [3.6], [4.5], [5.8], [8.0], and [24], respectively.
Thus, for each band, all reddened
SWIRE sources brighter than the completeness limit in Serpens would be detectable
by c2d-like observations and are identified as such in the resampled SWIRE catalog.  Most,
but not all, sources fainter than the completeness limit will not be detected by c2d-like 
observations.  We randomly select which sources to identify as detections, in a given band, 
in such as way as to reproduce the empirically determined shape of the completeness function.
This {\it{resampling}} process is performed for each band,
resulting in those sources fainter than the completeness limits in some bands being detected or
not detected with the same probabilities as those for similar sources in Serpens.  The
photometric uncertainties of all sources in the resampled catalog of reddened SWIRE sources
are re-assigned uncertainties similar to those of Serpens sources with similar magnitudes.

Finally, each source in the resampled SWIRE catalog is re-classified, based on its 
degraded photometry, e.g. ``star'', ``YSOc'', ``GALc'', ....  
The magnitudes, colors, and classifications of sources in this
resampled SWIRE catalog are then directly comparable to those in our Serpens
catalog and may be used to estimate the population of extragalactic sources
satisfying various color and magnitude criteria.  At this level of the classification, the terms
YSOc and GALc imply {\it candidate} classification status.

\begin{figure}
\plotfiddle{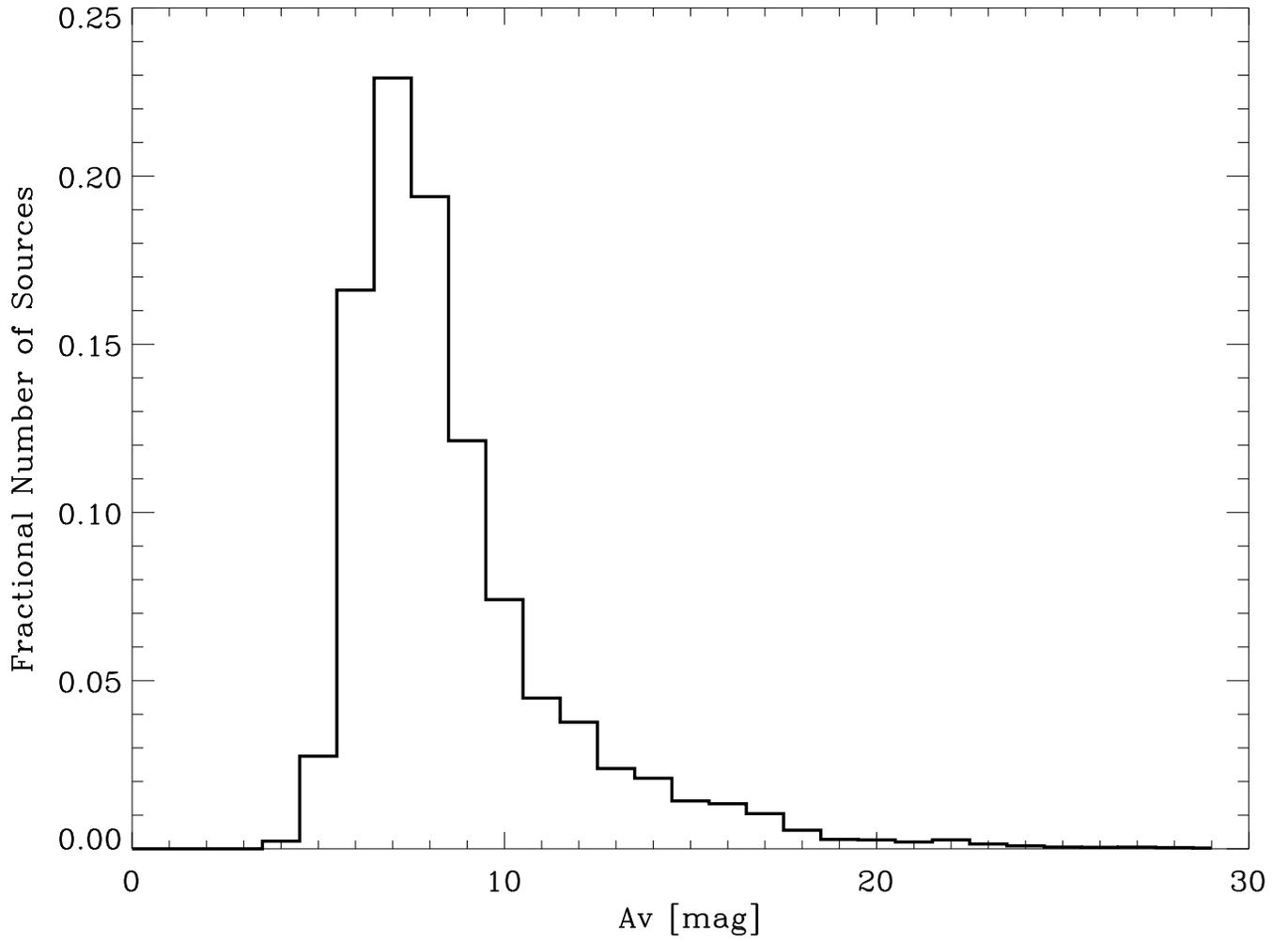}{7.0in}{90}{75}{75}{290}{0}
\figcaption{\label{histAV}
The distribution of visual extinctions found toward the roughly 50,000 sources classified
as stars in our Serpens observations.}
\end{figure}

\subsection{Classification Based on Color and Magnitude}\label{class}

In Paper I we described a simple set of criteria that 
basically categorized all objects that were faint and red in several combinations of
IRAC and MIPS colors as likely to be galaxies (after removal of normal reddened stars).
In our new classification we have extended this concept to include the color and magnitude spaces
in Figure \ref{cm_all} together with
several additional criteria to compute a proxy for the probability that a source
is a YSO or a background galaxy.  
Figure \ref{cm_all} shows a collection of three color-magnitude diagrams and one color-color
diagram used to classify
the sources found in our 3.6 to 70\micron\ survey of the Serpens Cloud that had S/N $\ge$ 3 in
all the Spitzer bands between 3 and 24\micron\ and that were not classified as reddened stellar photospheres.  
In addition to the Serpens sources shown in the left panels, the comparable set of sources from
the full-sensitivity SWIRE catalog are shown in the center panels, and the sources remaining in the
extincted/sensitivity-resampled version of the SWIRE catalog described above are shown in the right panels.
The exact details of our classification scheme are described
in Appendix A.  Basically we form the product of individual probabilities
from each of the three color-magnitude diagrams in Figure \ref{cm_all} and then use additional
factors to modify
that total ``probability'' based on source properties such as: its K - [4.5] color, whether it was found to be
extended in either of the shorter IRAC bands, and whether its flux density is above or
below some empirically determined limits in several critical bands.  Table \ref{prob_table}
summarizes the criteria used for this class separation.  The cutoffs in each of the
color-magnitude diagrams and the final probability threshold to separate YSO's from
extragalactic objects were chosen: (1) to provide a nearly complete elimination
of all SWIRE objects from the YSO class, and (2) to maximize the number of YSO's selected in
Serpens consistent with visual inspection of the images to eliminate obvious extragalactic
objects (such as a previously uncatalogued obvious spiral galaxy at RA = 18h 29m 57.4s , Dec = +00\degree\ 31' 41" J2000 ).

The cutoffs in color-magnitude space were constructed as smooth, exponentially decaying probabilities
around the dashed lines in each of the three diagrams.  Sources far below the lines were
assigned a high probability of being extra-galactic contamination with a smoothly decreasing
probability to low levels well above the lines.   In the case of the [24] versus [8.0]-[24]
relation, the probability dropped off radially away from the center of the elliptical
segment shown in the figure.  After inclusion of these three color-magnitude criteria, we
added the additional criteria listed in Table \ref{prob_table}.  These included: (1) a factor
dependent on the K - [4.5] color, $Prob/(K-[4.5])$, to reflect the higher probability of a 
source being an extra-galactic (GALc) contaminant if it is bluer in that color, (2) a higher GALc probability for
sources that are extended at either 3.6 or 4.8\micron\ where our survey had the best sensitivity
and highest spatial resolution, (3) a decrease in GALc probability for sources with a 70\micron\
flux density above 400 mJy, empirically determined from examination of the SWIRE data, and (4)
identification as ``extragalactic'' for {\bf any} source fainter than [24] = 10.0.
Again, we emphasize that these criteria are based only on the empirical approach of trying to
characterize the SWIRE population in color-magnitude space as precisely as possible, not on
any kind of modeling of the energy distributions.

Figure \ref{probcut} graphically shows the division between YSO's and likely extragalactic
contaminants.  The number counts versus our ``probability'' are shown for both the Serpens
cloud and for the resampled SWIRE catalog (normalized to the Serpens area).  This illustrates
how cleanly the objects in the SWIRE catalog are identified by this probability criterion.
In the Serpens sample there are clearly two well-separated groups of objects plus a tail
of intermediate probability objects that we have mostly classified as YSO's since no such tail
is apparent in the SWIRE sample.

Our choice of the exact cut between ``YSO'' and ``XGal'' in  Figure \ref{probcut} is somewhat
arbitrary because of the low level tail of objects in Serpens in the area of
log(probability) $\sim$ -1.5.
Since the area of sky included in our SWIRE sample is more than six times as large as the mapped area
of Serpens, we chose our final ``probability'' cut, $log(P) \le -1.47$, to allow two objects 
from the full (not resampled) SWIRE catalog into the ``YSO'' classification
bin.  Thus, aside from the vagaries of small-number statistics, we expect of order 0 -- 1
extragalactic interlopers in our list of Serpens YSO's.  The right panels in Figure \ref{cm_all},
which use the resampled version of the SWIRE catalog,
show that the effects of sensitivity and, especially, the extinction in Serpens make our cutoff
limits particularly conservative in terms of likelihood of misclassification.  
We examined all the YSO candidates chosen with these criteria both in terms of the quality of
the photometry and their appearance in the images.  A significant number, $\sim$ 50 candidates
were discarded because of the poor quality of the bandfilling process at 24\micron\ due to
contamination by a nearby brighter source or because of their appearance in the images.  It is
possible that a small number of these discarded candidates are, in fact, true YSO's in the
Serpens Cloud.  We also manually classified one source as ``YSO'' (\# 75 in Table \ref{yso-table}) that may be the
exciting source for an HH-like outflow in Cluster B (see also discussion of this region
by \citet{harv07}), but which was not so classified because of
its extended structure in the IRAC bands.   Table \ref{yso-table} lists the 235 YSO's that
resulted from this selection process.  

Another test of the success of our separation of background contaminants is to simply plot the locations
of the YSO's and likely galaxies on the sky.  Figure \ref{overlay} shows such a plot in addition to a 
plot of visual extinction discussed later.  There is
clearly a very uniform distribution of background contaminants and a quite clustered distribution
of YSO's (see further discussion in \S\ref{cluster}).  

\begin{figure}
\plotfiddle{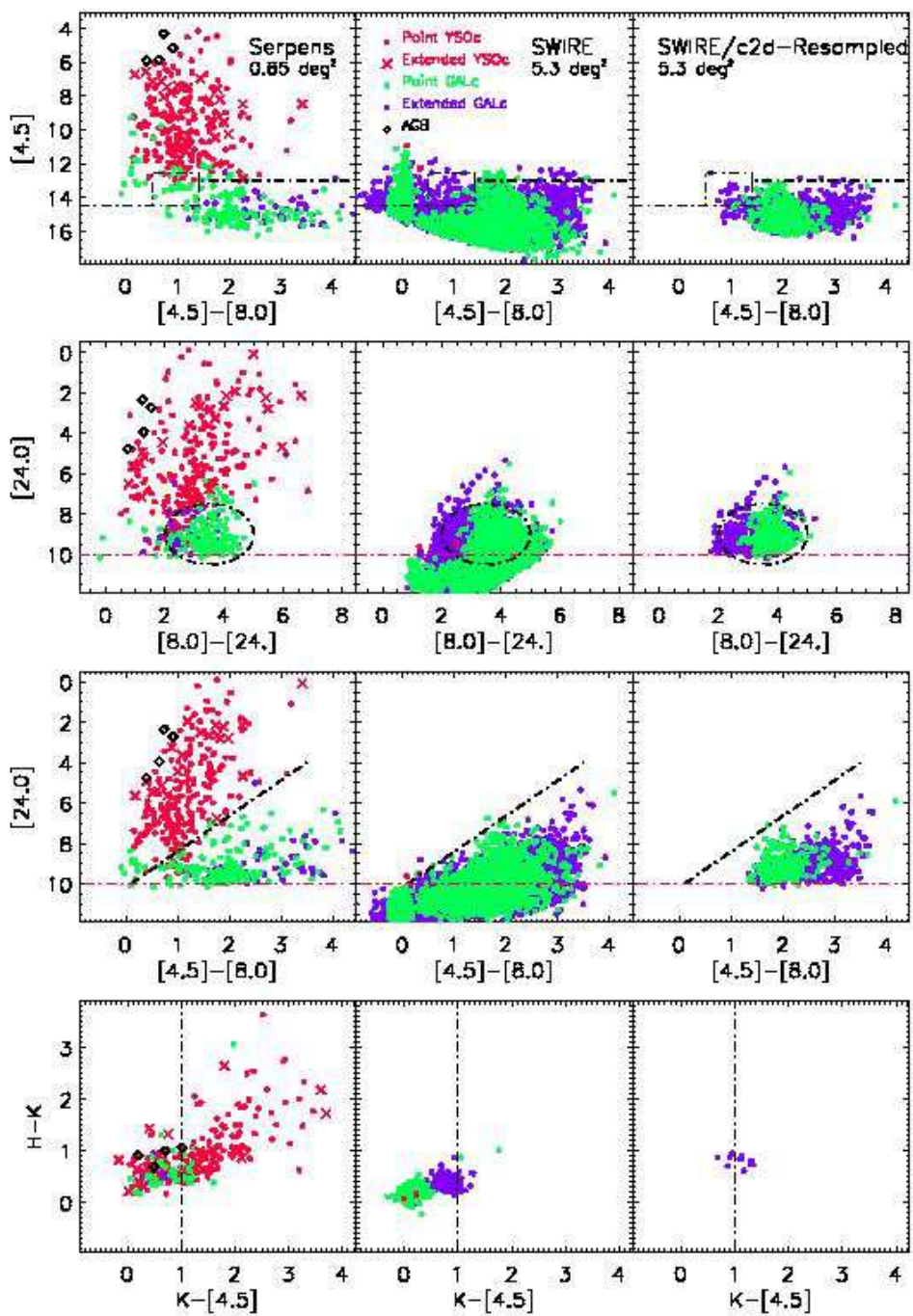}{7.0in}{0}{72}{72}{-220}{-10}
\figcaption{\label{cm_all}
Color-magnitude and color-color diagrams for the Serpens Cloud (left), full SWIRE (center), and
trimmed SWIRE regions.
The black dashed lines show the ``fuzzy'' color-magnitude cuts that define the YSO candidate criterion
in the various color-magnitude spaces. The red dashed lines show hard limits, fainter than which
objects are excluded from the YSO category.}
\end{figure}

\begin{figure}
\plotfiddle{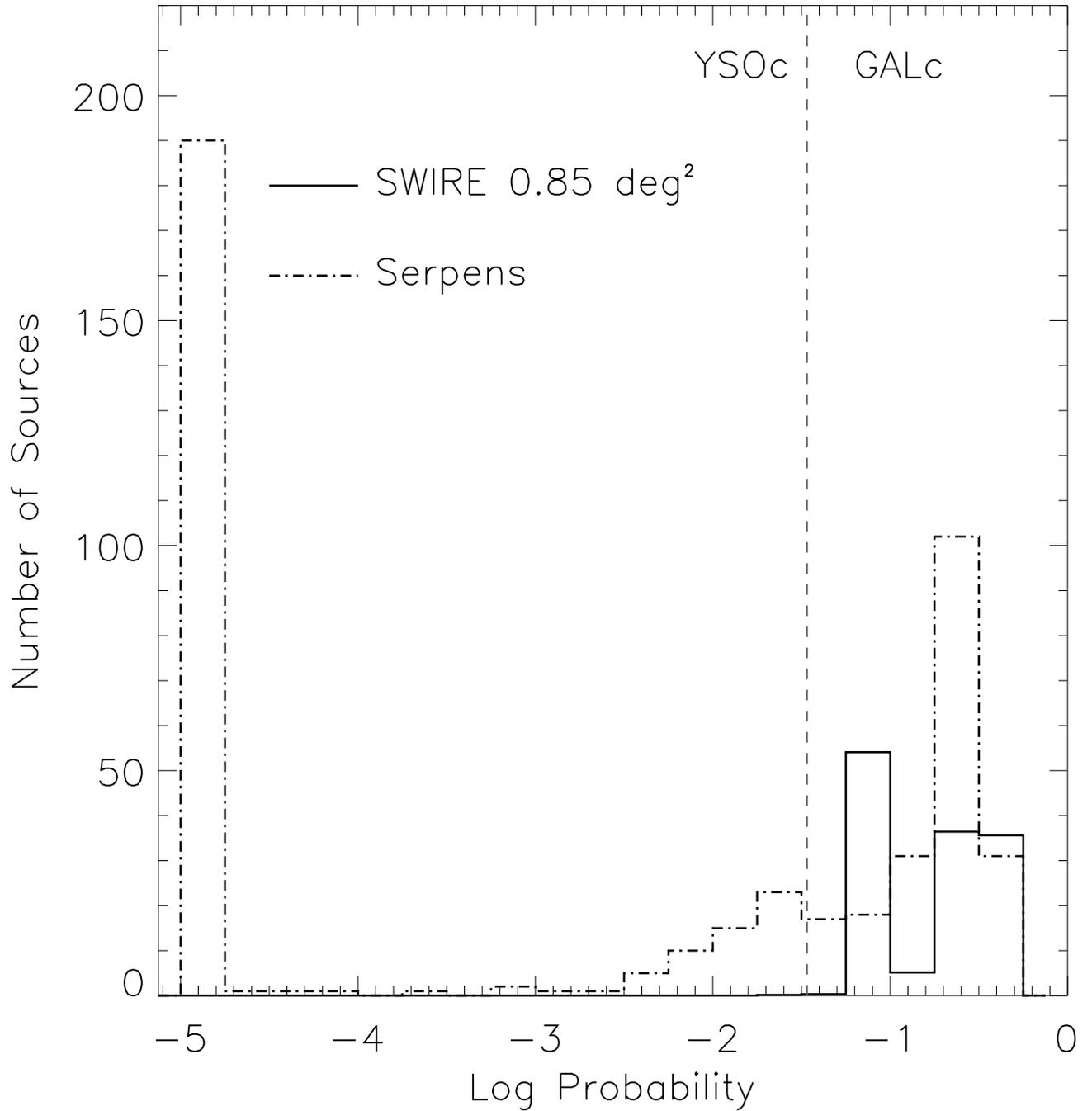}{7.0in}{0}{85}{85}{-260}{-90}
\figcaption{\label{probcut}
Plot of the number of sources versus probability (of being a background contaminant) for the Serpens cloud and
for the trimmed SWIRE catalog described in the text.  The vertical dashed line shows the separation chosen
for YSO's versus extra-galactic candidates. For both samples, only sources with detections in all four IRAC bands
are plotted to keep the number counts of contaminants on scale!}
\end{figure}

\begin{figure}
\plotfiddle{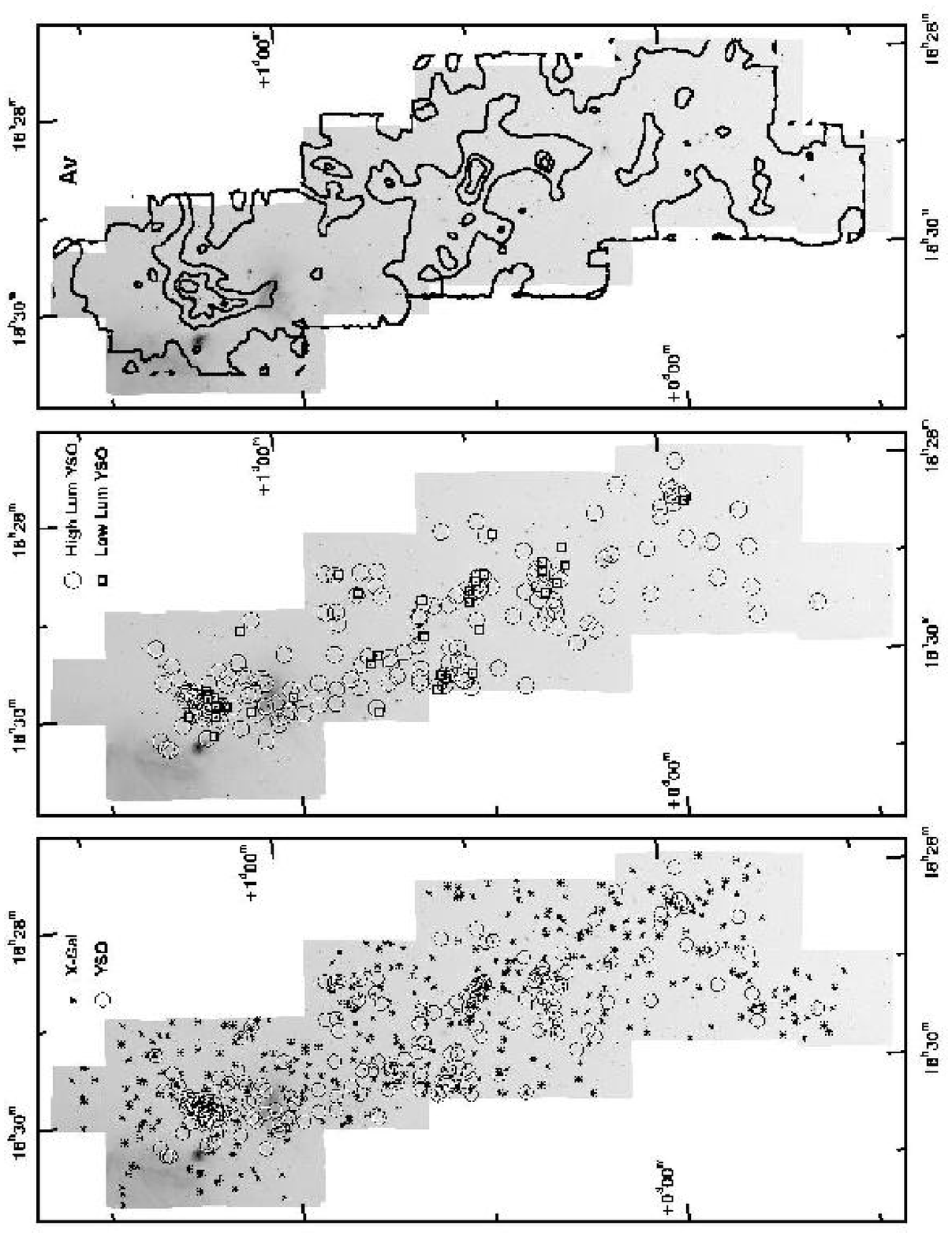}{7.5in}{0}{70}{70}{-230}{20}
\figcaption{\label{overlay}
Left panel: image of the entire mapped area of Serpens at 8.0\micron\ with the positions of YSO's plotted with
circles and likely extra-galactic background contaminants with asterisks; center panel: same image with YSO's of
luminosity $L < 2 \times 10^{-2}$\lsun\ plotted with boxes, and higher luminosity YSO's plotted with circles; right panel:
contours of visual extinction A$_v$ at levels of 5, 10, 20, 30 magnitudes as derived from fitting the energy distributions
of sources that were well-fit as reddened stellar photospheres in our data set.}
\end{figure}

There is one additional
kind of contaminant that is likely to appear in our data at a relatively low level, AGB
stars.  The ISO observations by, for example, \citet{iso99a} and \citet{iso99b} show that the range of brightnesses of AGB stars in the
LMC covers a span equivalent roughly to $8 < [8.0] < 12$, with colors generally equivalent to
$[4.5] - [8.0] < 1$.  Very compact proto-planetary nebulae are generally redder, but
even brighter at 8\micron\ (see e.g. Hora et al. 1996).  For typical galactic AGB stars, between 5 and 15 kpc from the sun,
this would imply $3 < [8.0] < 9$.   The off-cloud fields (when normalized to the same area as the Serpens data) 
provide the best handle on the degree of contamination from AGB stars; in Paper I we saw one object classified
as a YSO candidate in the off-cloud panel of Figure 9 with [8.0] $<$ 9.  Therefore, based on
the ratio of areas mapped in Serpens versus the off-cloud observations, we expect the number of AGB
stars contaminating the YSO candidate list in the Serpens cloud to be of order a half dozen.
As a further check, we have examined the entire set of off-cloud fields for such objects.  In this
combined area of 0.58 deg$^2$, there are only 3 such YSO candidates.   Finally, \citet{merin07} have classified
four AGB stars by their Spitzer-IRS spectra in the Serpens cloud; these are identified in the diagrams of
Figure \ref{cm_all} by black open diamonds and obviously are not included in our final list of high-probability
YSO's.  Statistically then, it is possible that a couple of our brightest YSO's are in fact AGB stars.  In fact,
\citet{merin07} have obtained modest S/N optical spectra of a number of our YSO's.  They find five objects whose
estimated extinction seems inconsistent with their location in the Serpens cloud and which suggests they might be 
background objects at much larger distances with modest circumstellar shells like AGB stars.  We
have indicated these five objects in Table \ref{yso-table} with a footnote ``b''.

It is also interesting to ask the reverse question, to what extent have our criteria successfully
selected objects that were known YSO's from previous observations.  \citet{alcala07} have used
the same criteria to search for YSO's in the c2d data for Chamaeleon II.  They conclude that all but one of the
known YSO's in that region are identified.  In \S \ref{compare} we discuss the comparison of our
results with several previous studies of Serpens that searched for YSO's.  Our survey found
counterparts to all the previously known objects in our observed area that had $S/N > 5$ in those studies, but we
did not classify many of them as YSO's because of the lack of significant excess in the IRAC or
MIPS bands.  Most of these were indeed suggested only as candidate YSO's by the authors, so we do
not consider this fact to be a problem for our selection criteria.  We again emphasize the fact that
our selection criteria for youth are based solely on the presence of an infrared excess at some Spitzer
wavelength.

The total number of objects classified as YSO's in the IRAC/MIPS overlap area of the Serpens cloud is 235.
From the statistics of our classification of the SWIRE data we 
expect of order 1 $\pm$ 1 of these are likely to be galaxies.
On the other hand, as we discuss in \S \ref{lowlum}, some of
the faint red objects in Figure \ref{cm_all} (below the dashed line) may also be young sub-stellar objects.  
For example, we note the case of the young brown dwarf BD-Ser 1 found by \citet{lod02} that was not selected
by our criteria because of its relative faintness.
Of these 235, 198 were detected in at least the H and K$_s$ bands of the 2MASS survey at better than 7$\sigma$.
The number of YSO's in each of the four classes of the system suggested by \citet{lada87} and extended by
\citet{greene94} is: 39 Class I, 25 Class ``Flat'', 132 Class II, and 39 Class III, using the flux
densities available between 1 and 24\micron.
Because we require some infrared excess to be identified as a YSO, the Class III candidates were
necessarily selected by some measurable excess typically at the longer wavelengths, 8 or 24\micron.
An obvious corollary is that objects identified as young on the basis of other indicators, e.g.
X-ray emission, lithium abundance, but without excesses in the range of 1 -- 24\micron\ are not
selected with our criteria.

\subsection{YSO's Selected By MIPS}

\citet{harv07} found 250 YSO candidates in the entire area mapped by MIPS at 24\micron\ in Serpens,
an area of 1.8 deg$^2$; 51 of these are outside the IRAC/MIPS overlap area and are listed in Table
\ref{mipsyso_table}.  We can make a comparison of those statistics with the YSO counts here in
two ways.  First in the area covered by both IRAC and MIPS24 there are 197 objects that satisfy the
criteria of \citet{harv07}, i.e., K$_s < 14$, K$_s - [24] > 2$, $[24] < 10$ and 24\micron\ S/N $\ge 5$.  Of these, 184 satisfy
our more restrictive criteria in this study based on the combination of 2MASS, IRAC, and MIPS data, or
93\%.  We would classify the other 13 as likely background galaxies.  Secondly, of the 235 YSO's found
in this study, 200 have sufficient data to be classifiable by the ``mips only'' criteria above, but
only 167 or 84\% actually meet the mips-only YSO criteria.  In other words, 33 objects have been classified as
high quality YSO's in this paper in the MIPS/IRAC overlap region that did not meet the criteria
based only on MIPS and 2MASS data.  If these ratios can be extrapolated to the larger area covered only
by MIPS, then we would expect \citet{harv07} to have missed 16\% (8 or 9) of the YSO's but to have included
7\% (3 or 4) that would not meet our combined IRAC/MIPS criteria.  With these corrections we might
have expected to find $235 + 56 = 291$ YSO's in the entire 1.8 deg$^2$ area covered by MIPS if we had matching IRAC
observations.  Finally, in light of our earlier discussion of AGB contaminants, it is possible that 6 -- 9 of
these YSO's would actually be found to be background AGB stars.

\section{Search for Variability}\label{vary}

The fact that our data were taken in two epochs separated by 6 hours or more gives us the opportunity to
search for variability over that time scale.  \citet{rebull06} and \citet{harv07} have performed similar tests
for variability of the 24\micron\ emission from sources in the c2d observations of Perseus and Serpens and found
no reliable evidence for variability at that wavelength.  There is, however, substantial evidence
for short term variability in the near-infrared for YSO's.  We, therefore, performed a similar
investigation in the two shortest IRAC bands, 3.6 and 4.5\micron.   No clear evidence was found
at the level of $\pm$ 25\% for any sources in the field over the 6-hour time scale of our multi-epoch
observations.

\section{Comparison with Previous Studies of Serpens}\label{compare}
 
We have cross-correlated our source catalog with those from previous studies
of Serpens that searched for YSO's.   We chose three studies that covered much of Cluster A at
both near-IR, ISO, and X-ray wavelengths \citep{ec92,kaas04,preib03}, and one recent ISO study of Cluster B
\citep{djup06}.  Table
\ref{cross-corr} lists the sources from each of these previous studies and the best-matching
Spitzer source from our complete catalog.  In brief, we find good matches for essentially all the previous
IR-selected YSO's that had S/N $>$ 4 in the earlier studies and which were included in our mapped
area.  In detail, however, a number of YSO candidates from the earlier studies were not
classified as YSO's in our study.  The reasons for this are different for the various catalogs.
From the X-ray catalog of Preibisch \citep{preib03} in Cluster A, we only identified 17 of the 45
X-ray sources as YSO's on the basis of their infrared excesses.  The X-ray sources that were
not identified as YSO's included both many objects that were well-fitted as reddened stellar photospheres
(21 sources) and objects with some likely infrared excess but too little to fit our criteria aimed at
eliminating extra-galactic interlopers.  The situation with the infrared catalogs is somewhat different.
Examining our non-matches from the ground-based study of \citet{ec92}, we find that a large fraction
of their YSO's are classified as such by our criteria, but not all.  From the ISO surveys \citep{kaas04,djup06}
we typically identify $\sim$50--60\% as YSO's.  The ones that we do not classify as such are typically those with
low S/N in the ISO observations or where the amount of infrared excess was not large enough to satisfy our test
of whether the object could not be fitted as an extincted stellar photosphere as described by \citet{evans07}.

\section{Luminosities}\label{lowlum}

\subsection{The YSO Luminosity Function in Serpens}

One of the most important physical parameters for any star is its mass.  For pre-main-sequence
stars this is problematic, because the determination of mass depends on the placement
of the star on an HR diagram {\it and} the use of model evolutionary tracks for which
there is significant uncertainty at the level of at least a factor of 1.5 -- 2 in the literature \citep{hw04,stass04}.  For young stars in
a cluster, a poor, but still useful proxy for mass is the stellar luminosity since it is
possible that most of the stars have formed more or less simultaneously.  With all these
caveats in mind, we display in Figure \ref{lum_func} the histogram of total luminosities for the
235 YSO's found in Serpens in our survey with the assumed distance of 260 pc.  We remind the reader that these objects were
selected specifically on the basis of infrared excess emission, so this list is limited to
YSO's that show substantial IR excess at least somewhere in the range of 3.5 - 70\micron.
For example, as we discussed in \S \ref{compare}, a comparison of our list of YSO's with the list of X-ray sources in Cluster A \citep{preib03}
showed that more than half the X-ray sources were not classified as YSO's by our infrared excess criteria.
In some sense then, it is likely that the IR-selected and x-ray-selected samples are complementary in
selecting less-evolved and more-evolved samples of YSO's respectively.
The luminosity function of our IR-excess-defined YSO sample peaks at
roughly  $2 \times 10^{-2}$ \lsun\ and drops off steeply below $10^{-2}$ \lsun.  This is interestingly equal to the luminosity
of objects at the hydrogen-burning limit of 0.075 \msun\ in the models of \citet{baraff02},
for objects with ages of $\sim$2 Myr, a common estimate for the age of the Serpens star forming event
\citep{kaas04,djup06}.
Of course an important question to ask is to what extent this luminosity distribution in Figure \ref{lum_func}
is influenced by selection effects.

We can estimate the selection effects in our sample from the statistics in the SWIRE samples discussed above.  
The completeness limits of the full catalog are nearly 2 magnitudes fainter than the trimmed
version, so we assume for simplicity that the full catalog is 100\% complete down to the faintest
magnitudes of interest for this test in the c2d Serpens data set.  
Figure \ref{completeness} shows the number counts as a function of ``luminosity'' (all sources
assumed to be at 260 pc) for three versions of our c2d-processed SWIRE catalog: 1) the full-depth catalog,
2) the catalog cut-off with c2d sensitivity limits but without added extinction, and 3) the catalog corrected
for both c2d sensitivity levels and extinction in the Serpens cloud.  
The ratio of the second to the first
version serves as a good proxy for a completeness function since it includes the effects of c2d sensitivity only.  
The lower panel of Figure \ref{completeness} then shows this ratio as our estimate of the completeness function
at the range of luminosities of Serpens YSO's observed in Figure \ref{lum_func}.
In Figure \ref{lum_func}, then, we also indicate how the YSO luminosity
function might be adjusted to account for this estimate of our survey completeness (with an
assumed completeness of 100\% at the bright end).  Although substantial
adjustments must be made to the number counts at fainter luminosities, the general conclusion remains intact
that the luminosity function peaks around $ few \times 10^{-2}$ \lsun\  and drops to both lower and higher
luminosities.  Interestingly, if our completeness estimates are valid, there may still be a substantial
population of IR excess sources down to luminosities of 10$^{-3}$ \lsun\ as was already suggested in Paper I.
For example, as noted earlier in \S \ref{class}, the young brown dwarf found by \citet{lod02} was not selected
as a YSO with our criteria because it was faint enough that its colors placed it into the color and
magnitude ranges where extra-galactic objects begin to be prominent.

\begin{figure}
\plotfiddle{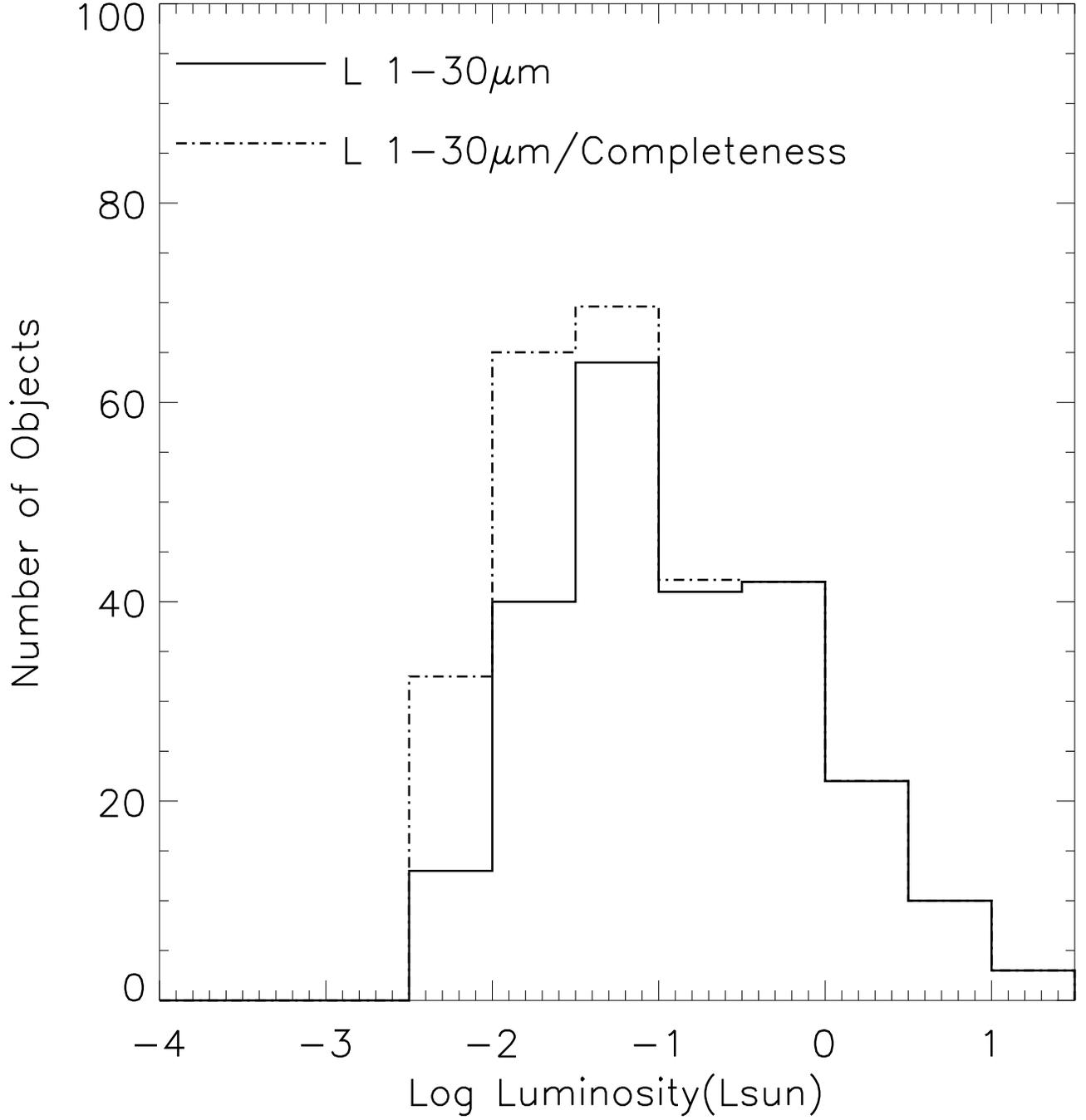}{7.0in}{0}{85}{85}{-260}{-90}
\figcaption{\label{lum_func}
Luminosity function for Serpens YSO's (solid) and estimate of correction for completeness effects
(dashed).}
\end{figure}

\begin{figure}
\plotfiddle{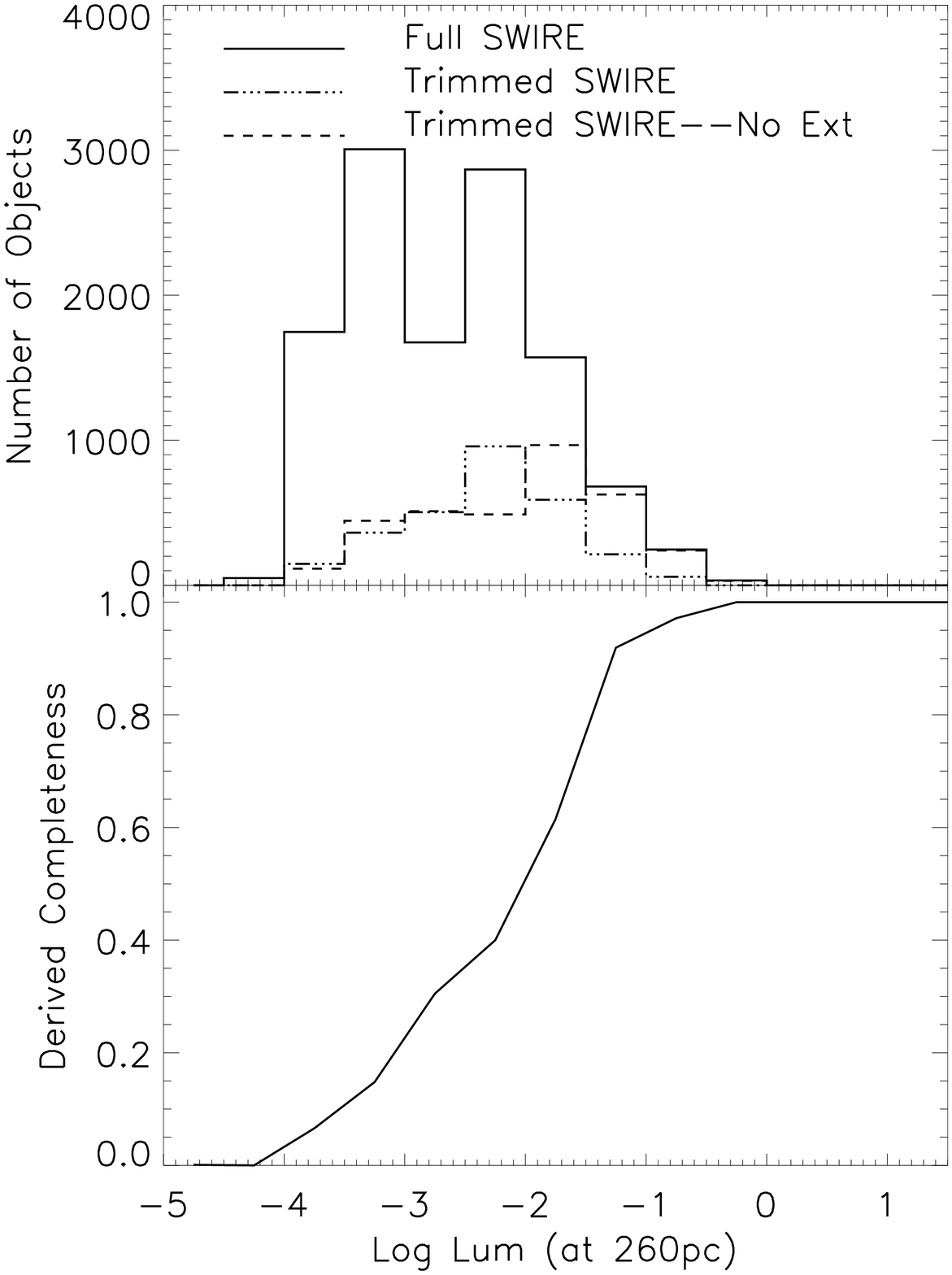}{7.0in}{0}{73}{73}{-230}{-30}
\figcaption{\label{completeness}
Upper panel -- number counts for the full SWIRE catalog as processed through the c2d pipeline for objects
detected in all four IRAC bands versus those for the ``trimmed'' version of the catalog with
completeness limits for each individual band comparable to those for c2d.  Lower panel -- the
ratio of the two number counts, i.e. completeness factor.}
\end{figure}

\subsection{The Lowest Luminosity Sources}

There are 37 YSO's with total luminosities (1 -- 70\micron) less than
$2 \times 10^{-2}$ \lsun, without any correction for selection effects in either our observations or
our selection criteria.
Figure \ref{overlay} presented in \S\ref{class} shows the spatial distribution of the
low (L $< 2 \times 10^{-2}$ \lsun) and ``high'' (L $> 2\times  10^{-2}$ \lsun) luminosity YSO's in Serpens.  There does not
appear to be any significant difference between these two distributions.
Likewise Figure \ref{alpha_hist} shows the distribution of spectral slopes, $\alpha$, for the low luminosity 
sample relative to two higher luminosity samples, and Table \ref{alpha_lum_table} lists the average
and standard deviation for the spectral slopes for three luminosity samples.
Again, there is no obvious difference in the distributions within the statistical
uncertainties of the samples.  (See also discussion in \S \ref{color_color}).
These two facts suggest that the mechanisms and timing of formation of these two luminosity groups
may  not be very different.  Finally, Figure \ref{av_hist} shows the distribution of extinction for stars
nearby each YSO.  For this analysis we selected all objects within 80" of each YSO that were classified as ``star'' by the 
process described earlier.  We averaged the fitted extinction values for these stars;
the number of stars contributing to the average ranged from 5 to 48.  This figure shows that the lower luminosity
YSO's have at least as much typical extinction as the higher luminosity ones, a fact that would be unlikely
if they were mis-classified background extragalactic objects.

\begin{figure}
\plotfiddle{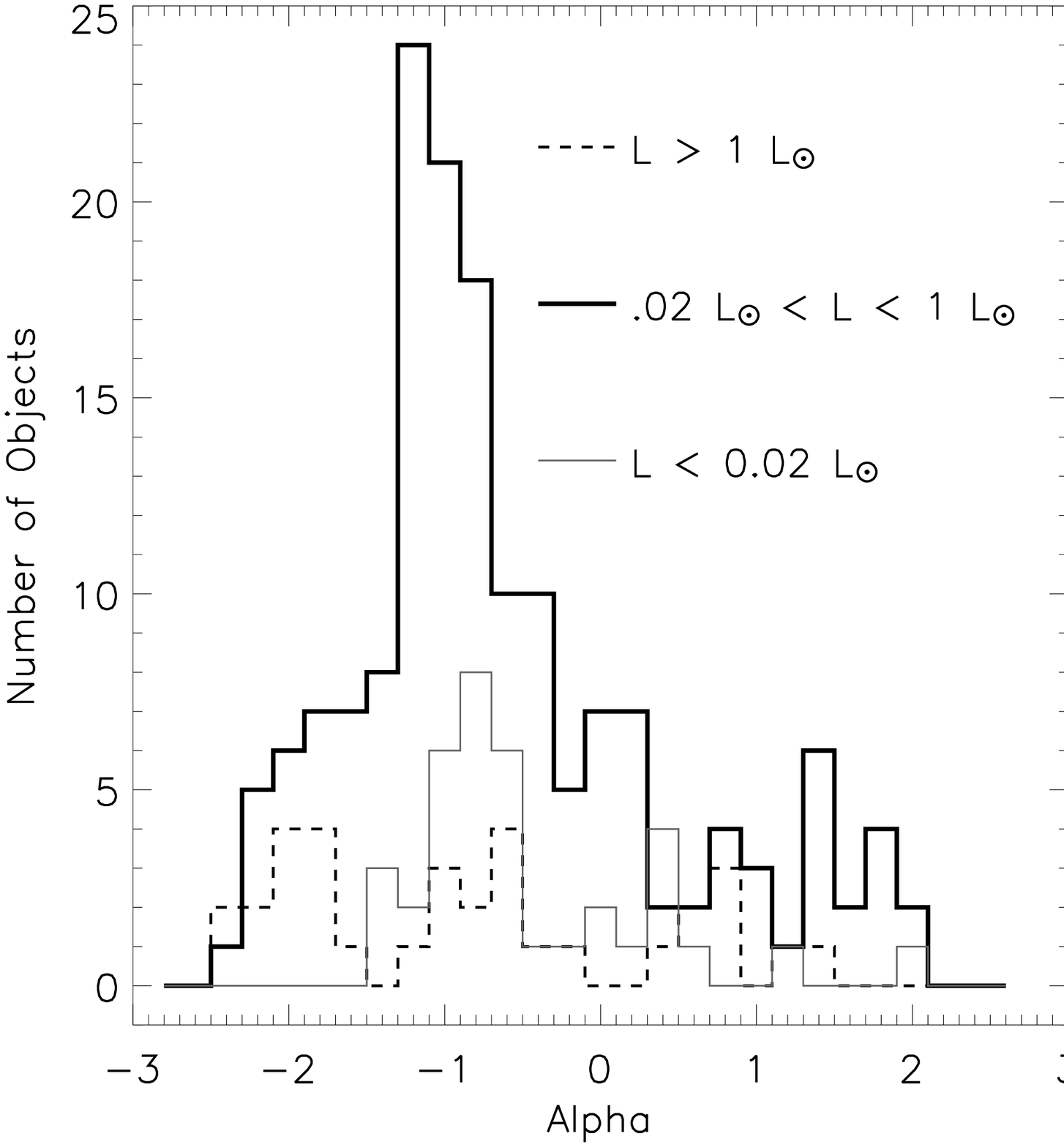}{7.0in}{0}{85}{85}{-260}{-90}
\figcaption{\label{alpha_hist}
Histogram of distribution of spectral slopes, ``alpha'', for three luminosity
``classes''
of YSO's.}
\end{figure}

\begin{figure}
\plotfiddle{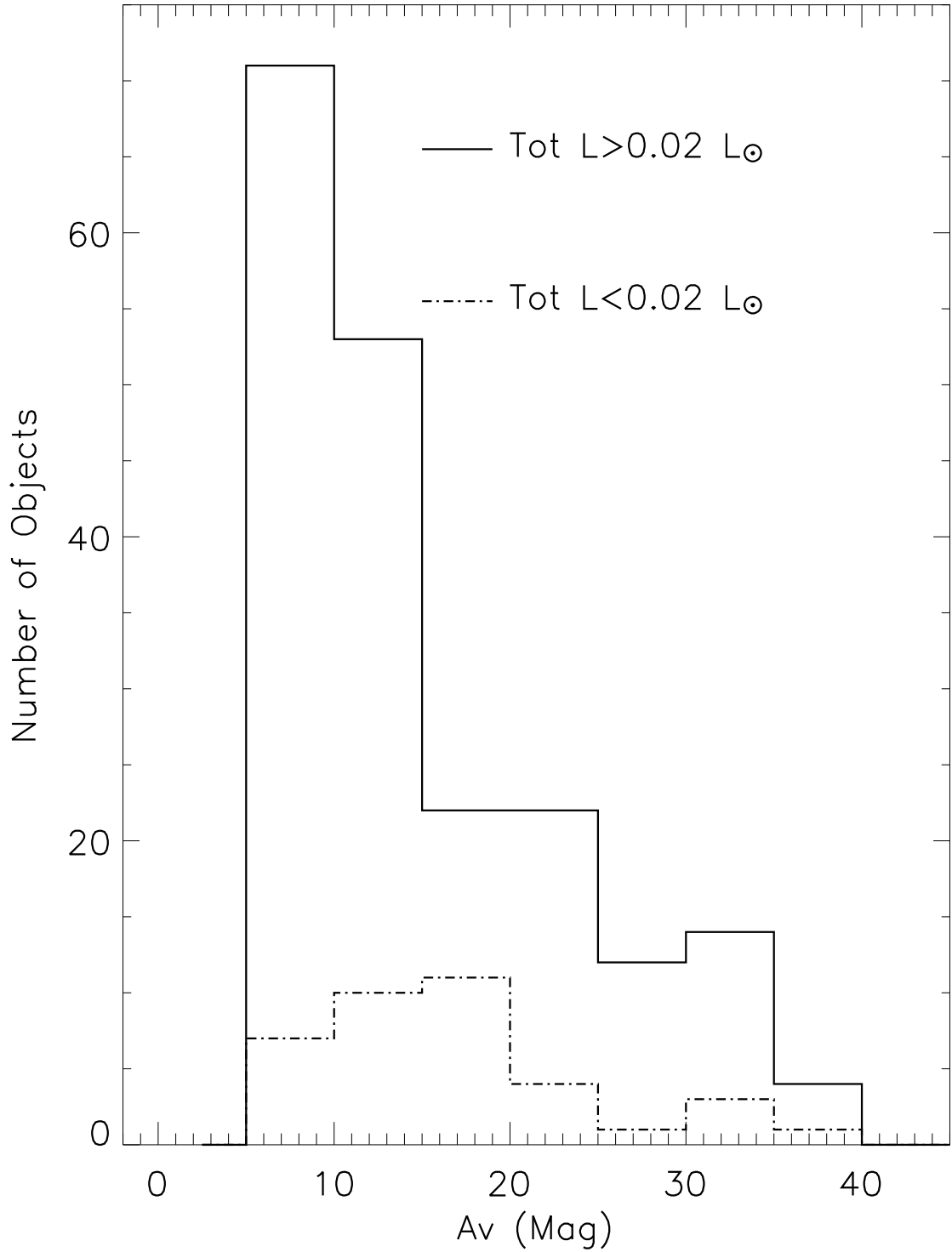}{7.0in}{0}{75}{75}{-260}{-15}
\figcaption{\label{av_hist}
Histogram of distribution of average extinction for stars within 80" of each
YSO for ``high''
and ``low'' luminosity YSO's (above or below $2 \times 10^{-2}$\lsun).}
\end{figure}

\section{Spatial Distribution of Star Formation}\label{cluster}

We have already noted in \S\ref{intro} that the Serpens star-forming region has been identified
as one displaying strong evidence of clustering of the youngest objects.  We also showed in Figure \ref{overlay} that
the spatial distribution of our high quality YSO candidates was highly clustered.  A number of authors
have attempted to quantify the degree of clustering as a function of spatial scale using the
two-point correlation function \citep{js00,enoch06,simon95,gl98,bate98}.  Typical results have found a steep slope, $\gamma \propto d^{-2}$ on small spatial 
scales implying a rapid decrease in clustering on scales out to $\sim$ 10000 AU, and a substantially
shallower slope, $\gamma \propto d^{-0.5}$ beyond that \citep{simon95,gl98}.  Figure \ref{cluster_w} shows the two-point correlation 
function, {\it W}, for
the YSO's in our Serpens catalog and the sample of objects classified as extra-galactic background sources.  The
Serpens samples are divided into two groups, those with spectral slopes, $\alpha$, that place them in the
Class I or ``Flat'' categories of \citet{greene94}, and those whose slopes imply Class II or III.
 Although our range of good sampling only extends down to
$\sim$ 1000 AU, equivalent to 4", our data suggest that the slope for separations under $10^4$ AU is of order 0.5,
and drops steeply for the more embedded and likely youngest  Class I and Flat-SED  objects beyond $10^4$ AU.  
The nominally more evolved objects in
Class II and III, exhibit a lower level of clustering and more uniform slope in the two-point correlation function.
These results are consistent with the appearance of the source distributions as shown, for example, in Figure 13 of Paper I.
As shown in Figure \ref{cluster_w}, the slope and magnitude of the correlation for our sample of background
extra-galactic objects is consistent with that found by, for example, \citet{maddox90} for a bright sample
of galaxies.

The total surface density of young stellar objects in a cluster is an indicator of the richness of
star formation in the cluster.  \citet{allen06} have computed surface densities for YSO's identified
by infrared excess in a sample of 10 young clusters.  They find typical peak surface densities of 500 -- 1000 pc$^{-2}$
and average values of order 100 -- 200 pc$^{-2}$.  The peak surface density in Cluster A of our YSO's identified
by IR excess is comparable to these, of order $10^3$ pc$^{-2}$, and a factor of about two less in Cluster B.  The average values
are about a factor of 4 less than the peak in both clusters.  In particular, if we define the cluster edges by the $A_v = 20$ contour
for comparison with other c2d clusters, 
we obtain the values in Table \ref{tabyso} for the number of stars per
solid angle and per square parsec. We have compared the values for clusters
A and B with those for the rest of the cloud and for the total cloud surveyed.
The surface densities of YSOs are 10 to 20 times higher in the clusters as
in the rest of the cloud.
As shown by \citet{allen06}, Cluster A in Serpens is
particularly striking in terms of the contrast between the peak surface density and how quickly, within 1 pc, the density
drops to very low values, $\sim few \times pc^{-2}$.
In terms of volume density, though, even Cluster A has a substantially lower volume density of star
formation than such rich clusters as the Trapezium (5000 pc$^{-3}$) or Mon R2 (9000 pc$^{-3}$) (Elmegreen et al. 2000).

Figure \ref{overlay} also shows the spatial distribution of visual extinction in the cloud as derived from
our data.  These extinction values were derived from the combination of 2MASS and Spitzer data for sources
that were well fit as reddened stellar photospheres as described in detail by \citet{evans07}.  As already noted in our previous studies of the IRAC
data alone \citep{harv06} and the MIPS data alone \citep{harv07}, there is clearly a striking correspondence between
the areas of high extinction and the densest clusters of YSO's, particularly those containing Class I and Class Flat
sources.

\begin{figure}
\plotfiddle{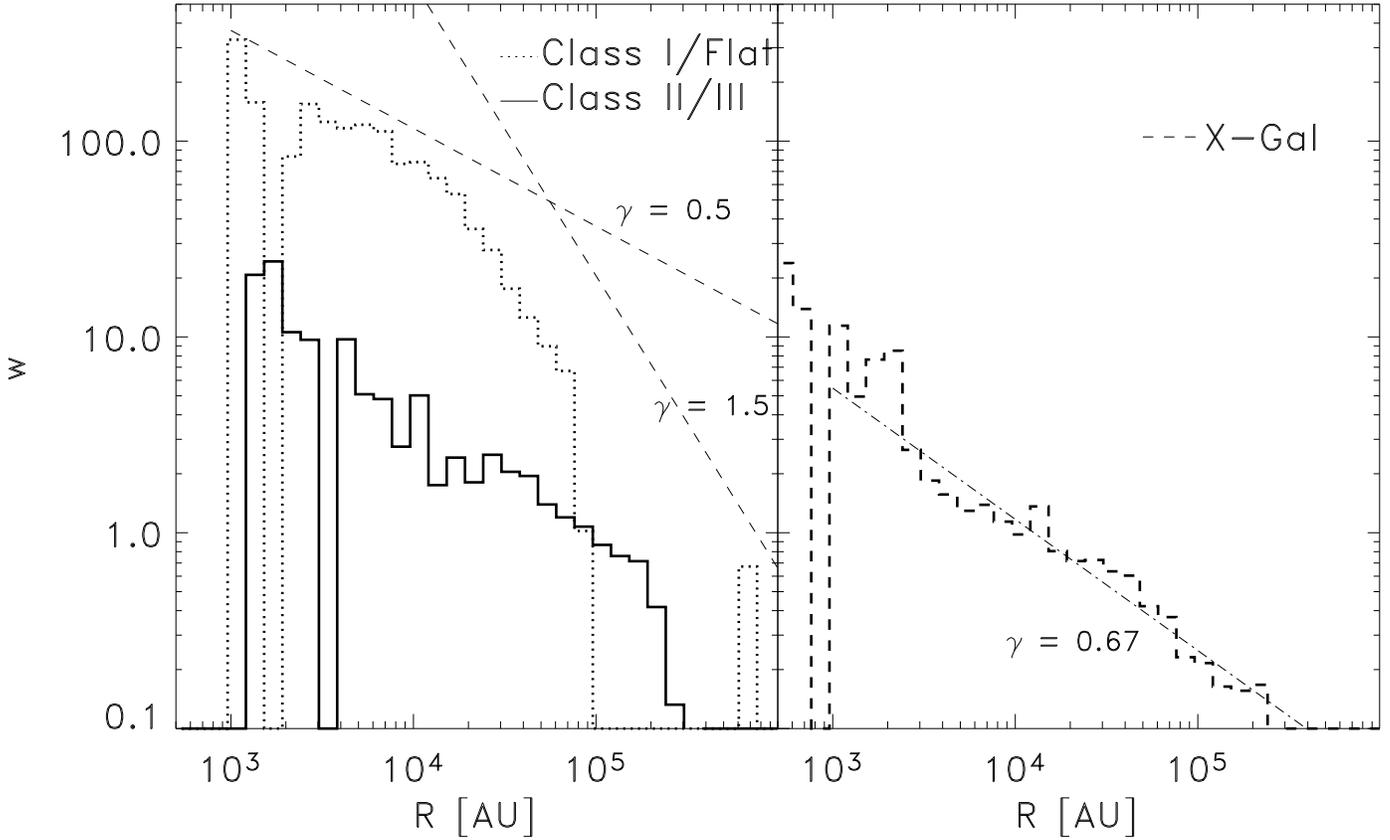}{7.0in}{90}{70}{70}{270}{-20}
\figcaption{\label{cluster_w}
Left panel: the two-point correlation function, {\it W}, for Serpens.  
The dotted histogram is for sources identified as Class I and ``Flat'' SED objects, i.e. $\alpha \ge -0.3$.
The solid histogram is for the Class II/III objects, i.e. $\alpha < -0.3$.  
The dashed lines show several
possible power law slopes for the portions of the YSO curves.
Right panel: the dashed histogram is for the objects
classified as likely background galaxies with S/N $> 7$ in all four IRAC bands.   
The dash-dot line shows the slope and
amplitude found by \citet{maddox90} for the bright end of the extra-galactic population, which is a quite
good fit to our estimate of the background contamination in our sample.}
\end{figure}

\section{Disk Properties}\label{disks}

The properties of the circumstellar dust surrounding young stars can be estimated from the overall energy
distributions of the objects in a variety of ways.  These properties include: the overall amount of dust, the
density distribution as a function of radial distance from the central star, the morphology of 
the circumstellar disks which probably represent the configuration of many of these circumstellar
distributions, and to some extent, the degree of evolution of the dust from typical interstellar dust
to larger and more chemically evolved grains likely to represent some stage in the formation of
planetary systems.  We discuss several approaches to classifying the dust emission
in this section, beginning with the use of infrared colors.

\subsection{Color-color diagrams}\label{color_color}

Numerous authors have shown how combinations of colors in the infrared can provide significant
diagnostic information on the configuration and total amount of circumstellar material around
young stars.  \citet{whitney03} modelled a wide range of disk and envelope configurations
and computed their appearance in a number of infrared color-color and color-magnitude diagrams 
particularly relevant to
Spitzer's wavelength range.  Similarly \citet{allen04} have published IRAC color-color diagrams
for another set of models of objects with circumstellar disks and/or envelopes.  
Recently, \citet{robit07} have computed a very large grid of models to extend the predictions from
those of \citet{whitney03} to a much larger parameter space.
Observationally, \citet{lada06} have used IRAC and MIPS colors from Spitzer
observations to classify the optical thickness of disks around pre-main-sequence stars in
IC 348, using the terms ``anemic disks'' and ``thick disks'' to point to objects roughly
in the Class III and Class II portions of the ubiquitous ``Class System'' \citep{lada87,greene94}.

Figure \ref{dsk_cccmall} shows three color-color diagrams and one color-magnitude diagram for
the 235 Serpens YSO's compared to the density of models from \citet{robit07} for a model cluster
at the distance of Serpens with brightnesses limited to those appropriate for our sensitivity
levels.  For two of the diagrams the figure also indicates the approximate areas where most of
the models fall for the three physical stages, I, II, and III, described by \citet{robit07} that roughly
correspond to the observational ``classes'' I$+$Flat, II, and III.
These diagrams display, first of all, that YSO's come in a wide range of colors and
brightnesses;  large areas of all the diagrams are occupied.
In a general way we find objects located in these diagrams where the models of \citet{whitney03},
\citet{allen04}, and \citet{robit07}  would predict for pre-main-sequence stars with a range of evolutionary
states as well as general agreement with the locations of young objects found in IC348 by
\citet{lada06}.
For example, consistent with our finding in paper I that the bulk of the YSO's are
class II objects, we see a strong concentration in the area around [5.8]-[8.0]= 0.8, [3.6]-[4.5]=0.5
where both Allen et al. and Whitney et al. predict such sources should be located.
Additionally, there is structure in the [8.0] vs. [3.6]-[8.0] color magnitude diagram for the distribution
of Serpens YSO's that is also evident in the density of models of \citet{robit07}.
There are, however, some interesting differences.  The ``Stage II'' area from \citet{robit07} includes
a smaller fraction of our nominal Class II objects than might be expected.
There is also a larger number of very red objects in several of the diagrams than is predicted
by the density of models in those areas, considering that the faintest grey levels correspond to
a probability equal to $10^{-4}$ of the maximum probability.  Some of this may be due to effects
of reddening, but it suggests that the models may also need some refinement, despite their
rough agreement with the distribution of observed colors.

These data can also be examined to see if there is any correlation between luminosity
and infrared excess, e.g. the [8.0] vs. [3.6]-[8.0] color magnitude diagram.  For example,
if higher luminosity young objects had higher levels of multiplicity (e.g. Lada 2006), then
it is possible this would be manifested in less massive disks with perhaps large inner holes.
Figure \ref{dsk_cccmall} shows no such correlation; in fact, if anything there appears to
be a lack of relatively blue colors for the brightest objects.  Figure \ref{alpha_hist} that
displays the distribution of spectral slopes and the accompanying Table \ref{alpha_lum_table}
also show no measurable effect within the scatter for a dependence of spectral slope on luminosity.

\begin{figure}
\plotfiddle{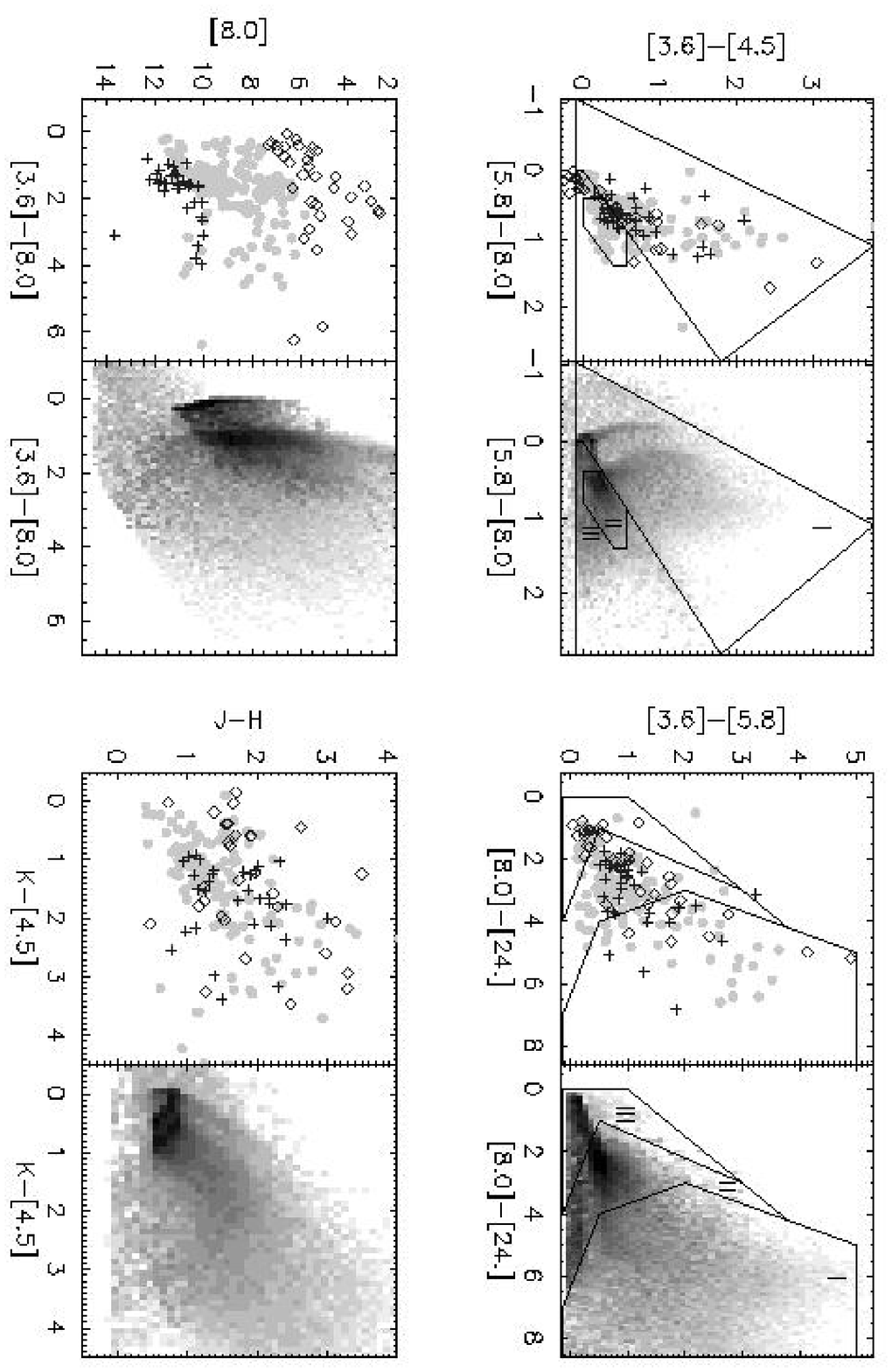}{7.0in}{180}{75}{75}{200}{580}
\figcaption{\label{dsk_cccmall}
Color-color and color-magnitude plots of the 235 Serpens YSO's compared with the distributions
from models of \citet{robit07}. The symbols are for the three different luminosity groups
discussed in the text: open diamonds, L $> 1$ \lsun; light filled circles, 0.02 \lsun $< L < 1$ \lsun;
plus signs, L $< 0.02$ \lsun. The model data were those for their model cluster, with a distance
of 260 pc for Serpens and with the c2d completeness limits.  Also shown are the rough areas from \citet{robit07}
occupied by mainly stage I, II, and III models.}
\end{figure}

\subsection{Spectral Energy Distribution Modeling}\label{sed}

In order to go beyond the general conclusions possible from color-color
and color-magnitude plots, we have modelled the large fraction of our
sources for which enough data are available to characterize them as
likely star-plus-disk objects.
For this analysis we have selected all Class II and Class III YSO's.
We expect these objects to have a reddened photosphere plus an infrared
excess coming from the circumstellar material, most likely in a disk configuration. For completeness
we show, without modeling, the SED's of the remaining YSO's. Figures \ref{sedI0} to \ref{sedIII1} show the
SEDs of all Class I, Class Flat, Class II and Class III sources. The open dots
are the observed fluxes and generally include 2MASS J, H, and K fluxes
followed by the four IRAC fluxes at 3.6, 4.5, 5.8 and 8.0 $\mu$m and
by the MIPS fluxes at 24 and 70 $\mu$m when available.  For the Class
IIs and IIIs, we characterize the emission from the disk by comparing
the energy distribution with that of a low-mass star. For this, we
have computed the extinction from the J-K colors assuming a K7
underlying star, and then we have dereddened the observed fluxes with
the interstellar extinction law of \citet{wein01} with R$_v = 5.5$ to fit the
dereddened SEDs to a NEXTGEN (Hauschildt et al. 1999) model of a K7
stellar photosphere. In the plots, the filled dots are the dereddened
fluxes and the grey line is the photospheric emission from the star.
Each SED is labeled with the identification number in Table \ref{yso-table}.
The parameters of the model fits are listed in Tables \ref{resultsII} and \ref{resultsIII}.

Two main assumptions are used in this approach. The first is that the
star is a low-mass star; this is reasonable given the fact that the
luminosity function discussed earlier peaks at $L \sim few \times 10^{-2}$\lsun.
In addition, Oliveira et al. (2007) have found that 30 of the 39 YSO's they
observed out of our sample had spectral types of K or M.
The second assumption is that there is no substantial excess in the
near infrared bands, where we consider all flux to be
photospheric. This assumption is obviously only correct for those
stars in the sample without strong on-going disk accretion. \citet{cieza05}
have demonstrated that near-IR excesses as short as J-band can
be seen in some actively accreting T Tauri stars, and they give an
observed range of J-band excesses of 0.1 -- 1.0 mag for their
sample of sources in Taurus. The effect of this excess on the
computed parameters implies that the stellar luminosities for the Class
II objects are upper limits and therefore the disk luminosities are
lower limits. More specifically, neglecting an excess flux in the J-band of
1.1 magnitude in one object would reduce its Ldisk/Lstar ratio by
12 \% and modify the disk emission SED, diminishing
its excess mostly at the shorter wavelengths.  That would
only happen, however, to the fraction of objects that are strongly accreting. Oliveira
et al. (2007) found strong H-$\alpha$ emission indicative of active
accretion in only 12 out of 39 Serpens YSO's observed. We ran a test removing
these objects from the final statistics and this did not change the overall
picture.

The dashed line in the SED plots is the median SED of T Tauri stars in
Taurus \citep{hartman05}  
normalized to the dereddened J-band flux of our SEDs. It represents
the typical SED of an optically thick accreting disk around a Classical
T Tauri star and is shown here to allow a qualitative estimation of
the presence of disk evolution and dust settling. Based on 
similarity to this median SED, we define as {\sl T} an SED which is
identical to it, within the errors, as {\sl L}, an SED with lower
fluxes at some wavelengths, and {\sl H} an object with larger
fluxes than the median T Tauri SED. We have labeled the objects with
these codes and use them to interpret the state of disk evolution. We
interpret the L-type objects as thin disks which could perhaps result from dust
grain growth and settling to the mid-plane, similar to the ``anemic
disks'' in \citet{lada06}. Objects 3 and 5 are good examples of
T-type objects, numbers 11 and 18 are L-type stars, and number 8 is a
H-type object. Also, we have labeled as {\sl LU} objects that have
photospheric fluxes up to around 8$\mu$m but then a sudden jump at
longer wavelengths to the levels of a T-Tauri disk. We believe these
objects may have large inner holes in their still optically thick
disks; number 19 is one example of these.

Out of the 171 sources in the sample that are Class II or Class III, 44 (26
\%) are T-type, so equivalent to a classical T Tauri star, 125 (73 \%)
are L-type, so show evidence of some degree of disk evolution, and
finally 2 (1 \%) show larger fluxes than the median SED of Taurus.
This latter category is generally the result of imperfect
fitting to the stellar photosphere due to lack of near-IR data and
will not be considered further.

The numbers above indicate that there are many young stars
in Serpens with disks that are virtually identical to those in 
Taurus. This points to a very similar origin of the disk structure
across star forming regions. Within the uncertainties in our modeling process discussed above,
however, we see a majority of L-type
objects (73 \%) in Serpens. This suggests that the disks in Serpens may be generally
more evolved or flattened that those in Taurus. Also, interestingly
the number of LU-type objects, or those with large inner holes, amounts
to 14 (11 \% of the L-type objects and only 6 \% of the total YSO
sample including Class I and F sources). This very small incidence
indicates that either the process producing these large inner holes is
a very fast transitional phase in the disk evolution or that only a
small number of systems undergo that phase. We examine this issue
below.  In order to put firmer limits on these numbers it will be necessary
to get optical spectral types for the stars and optical photometry in order
to produce more accurate photosphere/disk/extinction models.

The SEDs also allow a rough study of the disk properties across the
surveyed area. For this purpose we integrated the stellar and total
fluxes to calculate a ratio between the stellar and the disk fluxes.
The corresponding distribution is shown in Figure \ref{diskL}. Again, the solid
line is the whole Class II and III sample and the dotted line
corresponds to the T-type SEDs. To guide the eye, we have marked the
approximate regimes of these ratios measured in debris disks
(Ldisk/Lstar $<$ 0.02), passive disks (0.02 $<$ Ldisk/Lstar $<$ 0.08
Kenyon \& Hartmann 1987), and accretion disks (Ldisk/Lstar $>$ 0.1),
respectively. The figure illustrates well the large variety of disk
evolutionary phases that we observe in Serpens, as well as the large
majority of massive, accreting disks in the region.  

In order to study the disk evolution in the cloud, there is need for a
better characterization of the full sample with optical
spectroscopy. That work is in progress and will be published
elsewhere. Here we explore further the shapes of the SEDs to test
another evolutionary diagnostic of the disk status.  We have found that
the more commonly used evolutionary metrics, e.g. spectral slope, $\alpha$,
bolometric temperature, $T_{bol}$, or the Class system discussed earlier
do not seem to capture the full range of phenomena apparent in our
distribution of SED's for young stellar objects.  We, therefore, use
$\alpha_{\rm excess}$ and $\lambda_{\rm excess}$, two new
{\sl second-order} SED parameters presented in \citet{cieza06} that
will be discussed in more detail in a separate paper. In short,
$\lambda_{\rm excess}$ is the last wavelength where the
observed flux is photospheric and $\alpha_{\rm excess}$ is the slope
computed as $dlog(\lambda F_\lambda)/dlog(\lambda)$ starting from
$\lambda_{\rm excess}$. 
The first parameter gives us an indication of
how close the circumstellar matter is to the central object and the
latter one a measure of how optically thick it is. Given the
assumptions above for the fitting process, our values are upper limits
for $\lambda_{\rm excess}$, and correspondingly lower limits for
$\alpha_{\rm excess}$ for the Class IIs. We assume good stellar fits
for the Class III sources.  

Figure \ref{alpha_turn} shows both values for a sample of Classical T Tauri stars
(CTTs) from \citet{cieza05}, weak-lined T Tauri stars (wTTs) from
\citet{cieza06}, debris disks from \citet{chen05}, and our 
sample of YSOs in Serpens. \citet{cieza06}
found that $\lambda_{\rm excess}$ is well correlated with evolutionary phase
and also, as seen in Figure \ref{alpha_turn}, that $\alpha$ is observed over wider
ranges for later evolutionary phases.
This suggests a
large range of possible evolutionary paths for circumstellar
disks. Objects could either evolve to the lower right side of the
diagram, which could be interpreted as a sign of dust grain growth
and settling to the disk mid-plane; they could also move to the 
right and up, from creation of a large inner hole in their disks while
retaining a massive outer disk, and also, as this diagram shows,
all intermediate pathways in between both extremes.

\begin{figure}
\includegraphics[angle=0,width=15cm]{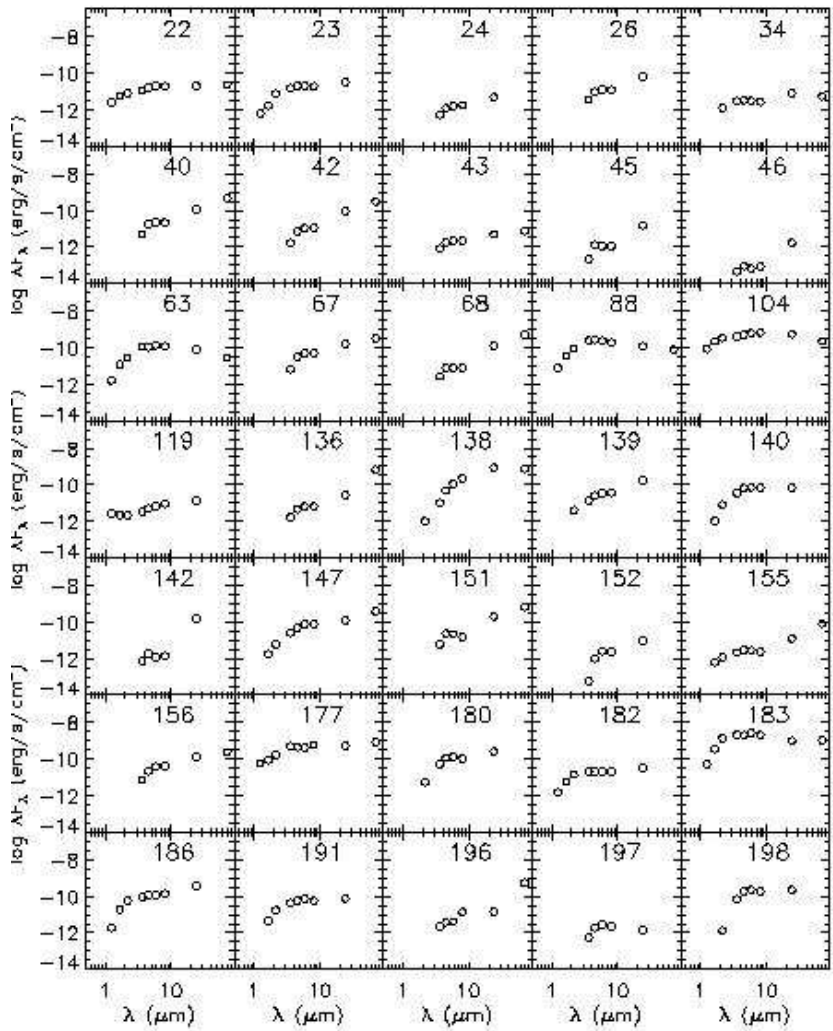}
\caption{\label{sedI0}Spectral Energy Distributions of the Class I sources in 
the sample.
The open dots signal the observed fluxes from J-band to MIPS-70 when
available. The label gives the index in Table \ref{yso-table}.}
\end{figure}

\clearpage
\begin{figure}
\plotfiddle{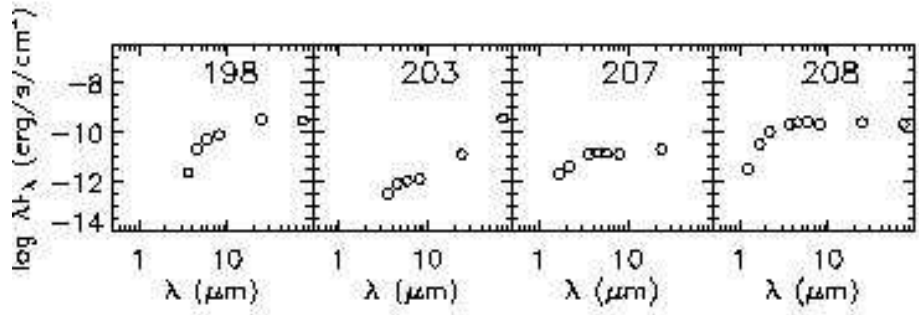}{4.0in}{0}{100}{100}{-180}{-400}
\caption{\label{sedI1}Spectral Energy Distributions of the Class I sources in 
the sample (continued). Symbols as in Figure \ref{sedI0}.}
\end{figure}
\clearpage

\begin{figure}
\includegraphics[width=15cm]{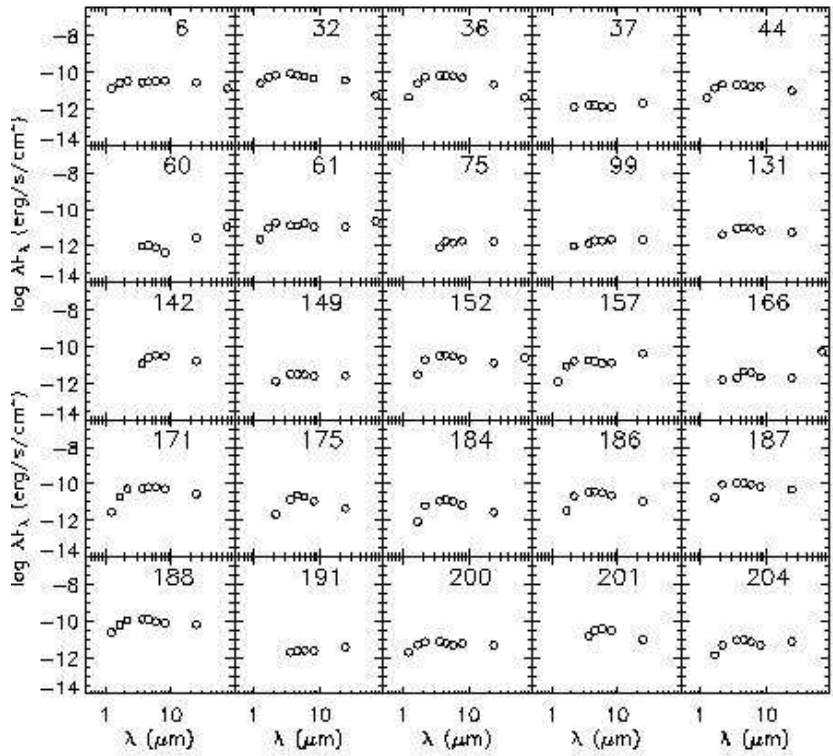}
\caption{\label{sedF0}Spectral Energy Distributions of the Flat sources in the sample. Symbols as in Figure \ref{sedI0}.}
\end{figure}

\begin{figure}
\includegraphics[width=15cm]{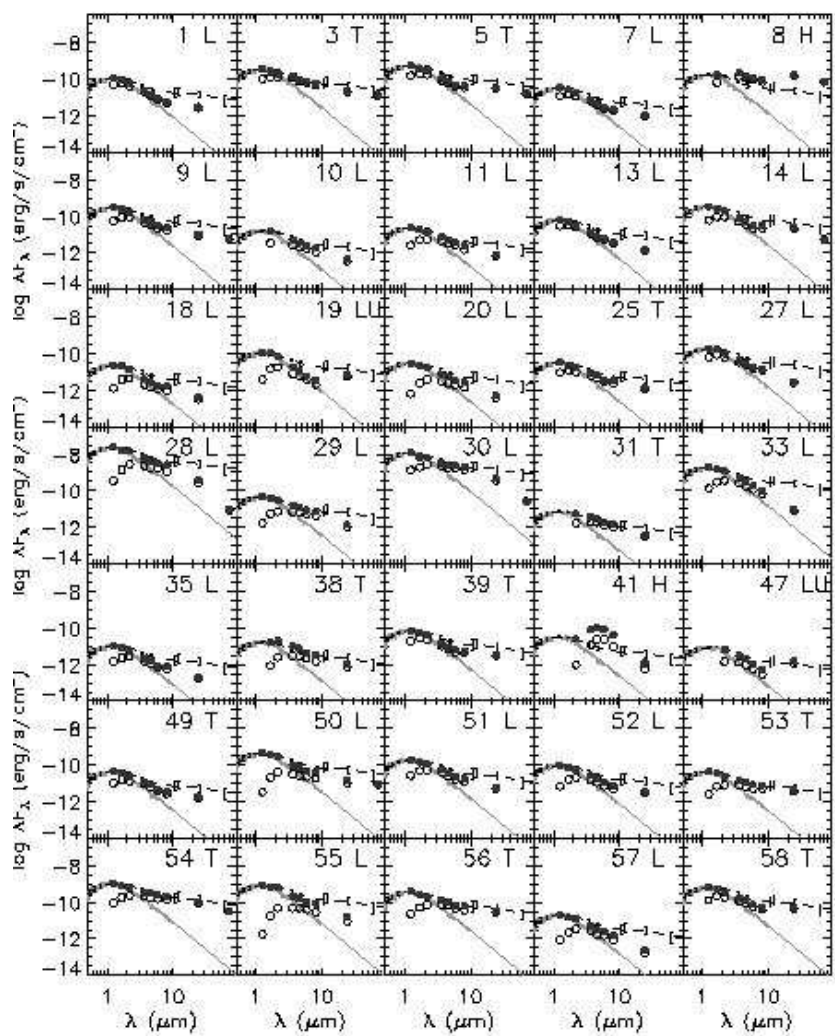}
\caption{\label{sedII0}Spectral Energy Distributions (SED) of the Class II sources in the
sample. The open and solid dots are the observed and dereddened fluxes
respectively. The grey line is the stellar model of a K7 star and the
dashed line is the median SED of the TTauri stars in Taurus by Hartmann
et al. (2005) normalized to the dereddened J-band flux for comparison.
See text for more information.}
\end{figure}

\begin{figure}
\includegraphics[width=15cm]{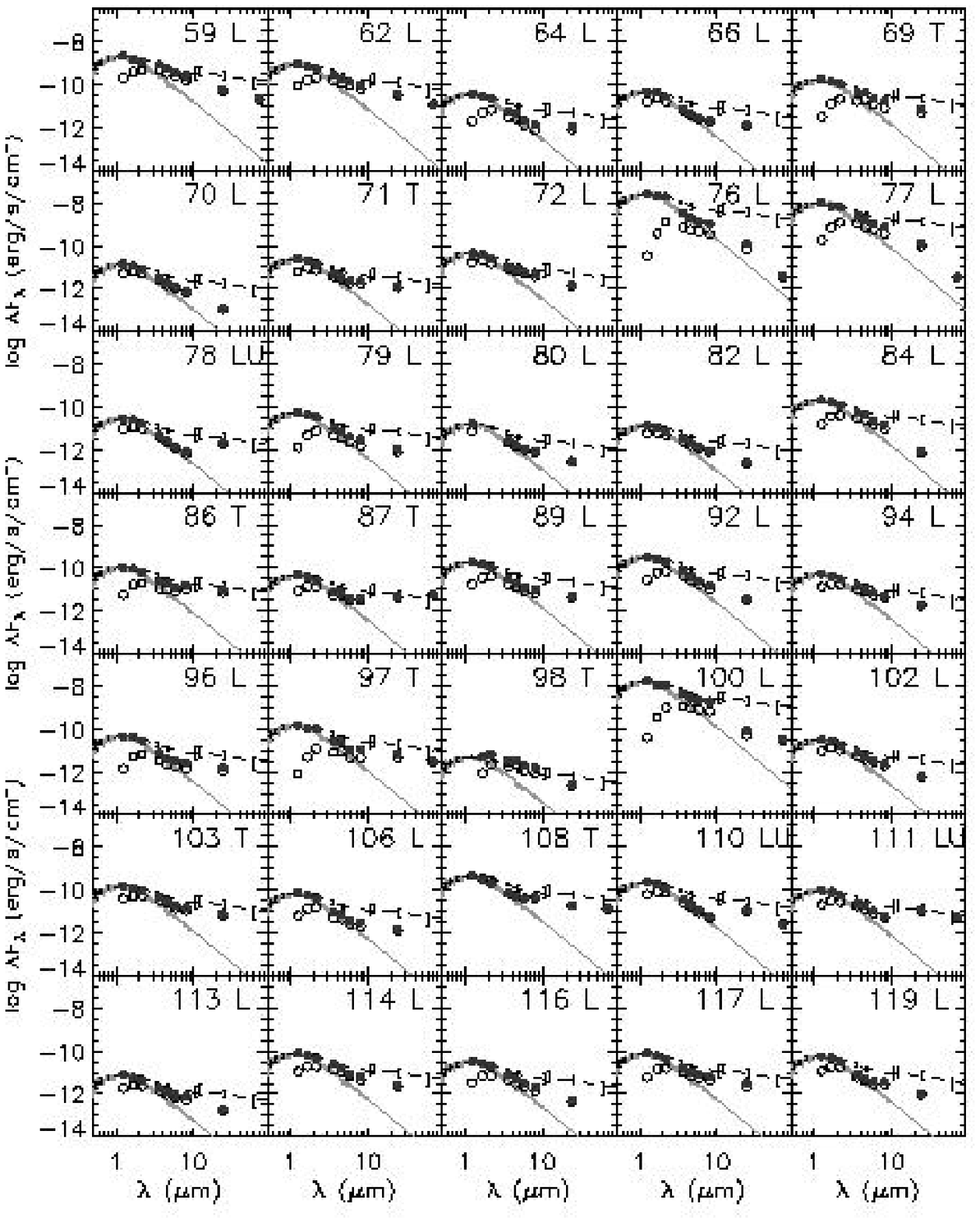}
\caption{\label{sedII1}Spectral Energy Distributions of the Class II sources in the
sample (continued).}
\end{figure}

\begin{figure}
\includegraphics[width=15cm]{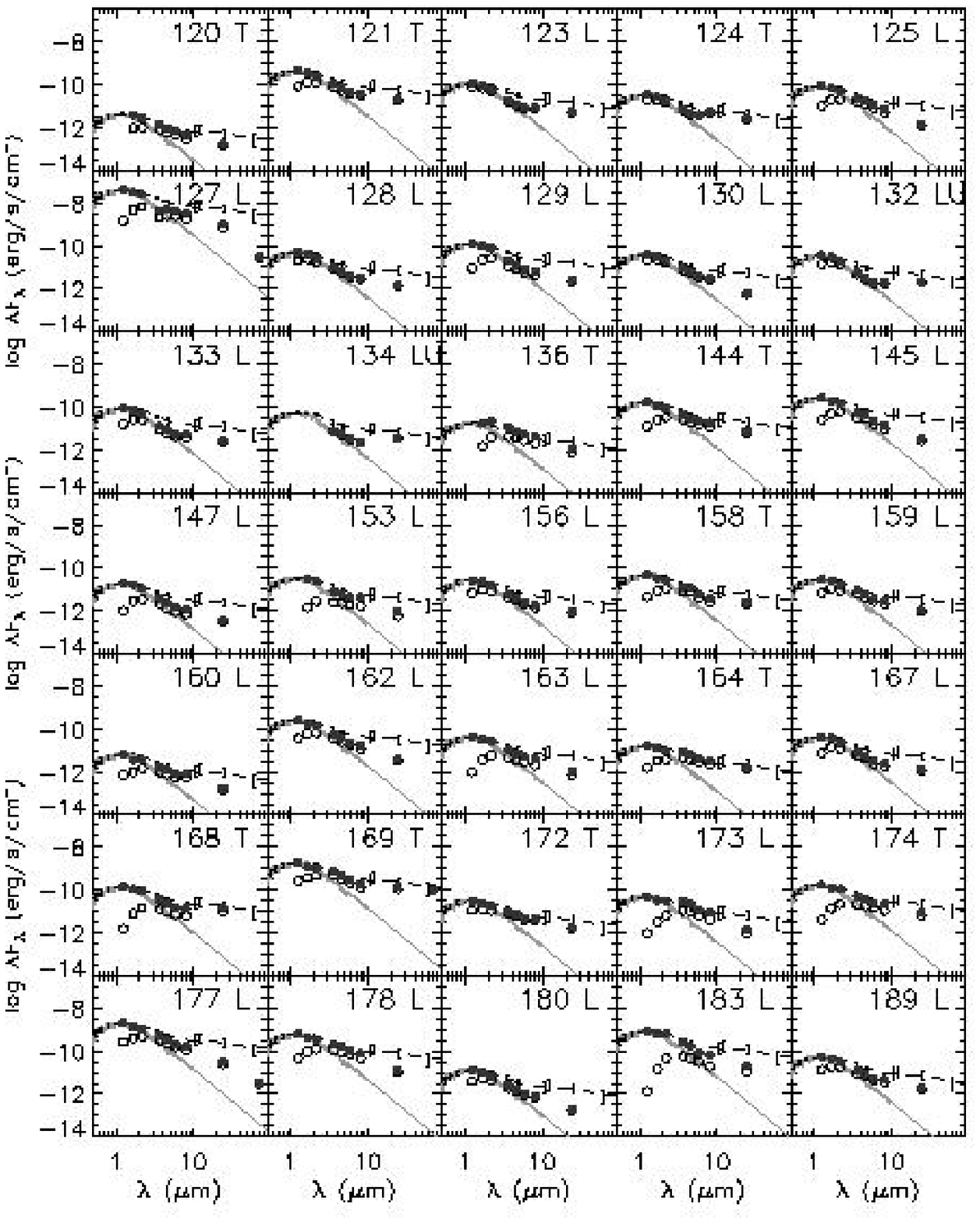}
\caption{\label{sedII2}Spectral Energy Distributions of the Class II sources in the
sample (continued).}
\end{figure}

\begin{figure}
\includegraphics[width=15cm]{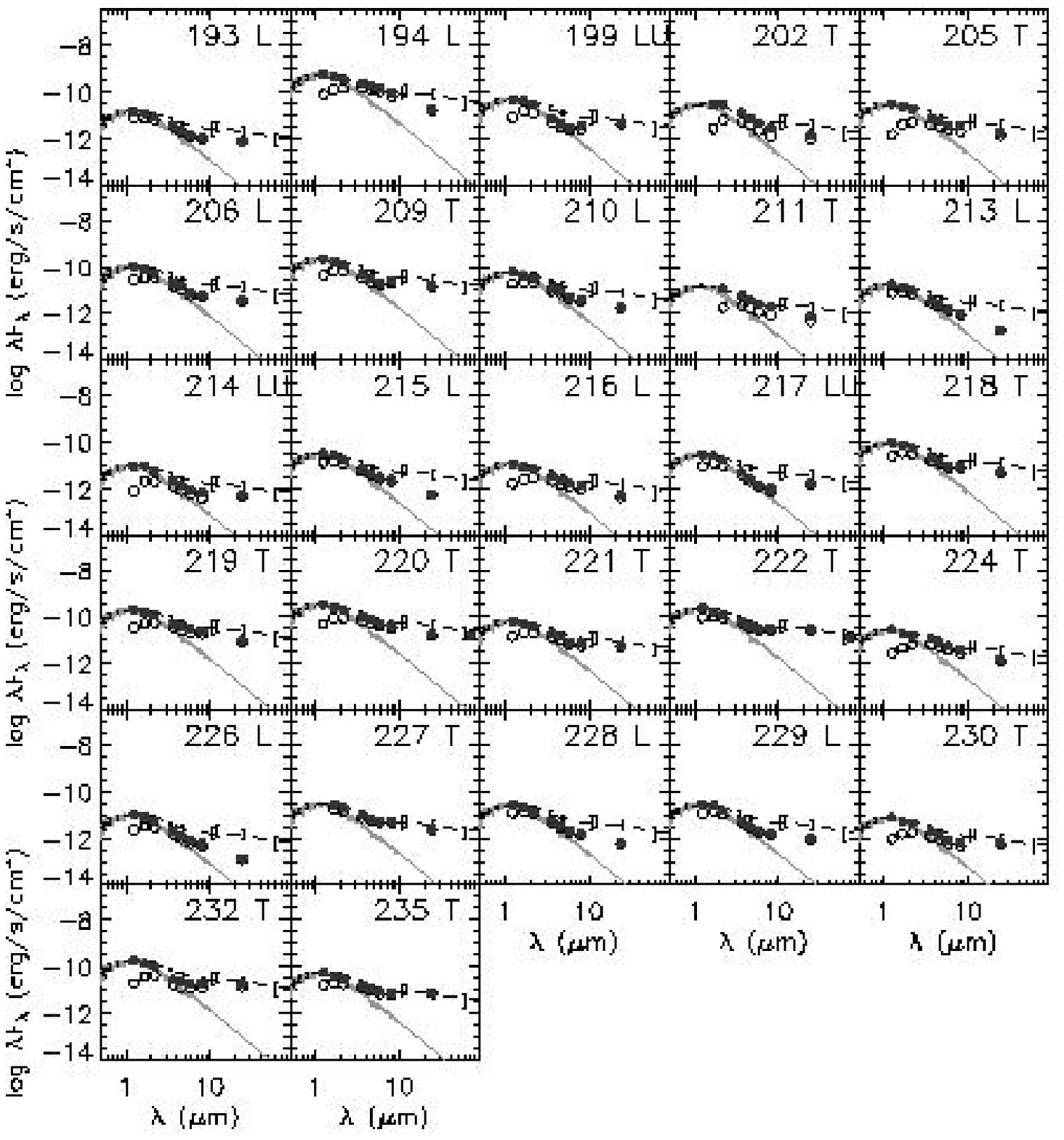}
\caption{\label{sedII3}Spectral Energy Distributions of the Class II sources in the
sample (continued).}
\end{figure}

\begin{figure}
\includegraphics[width=15cm]{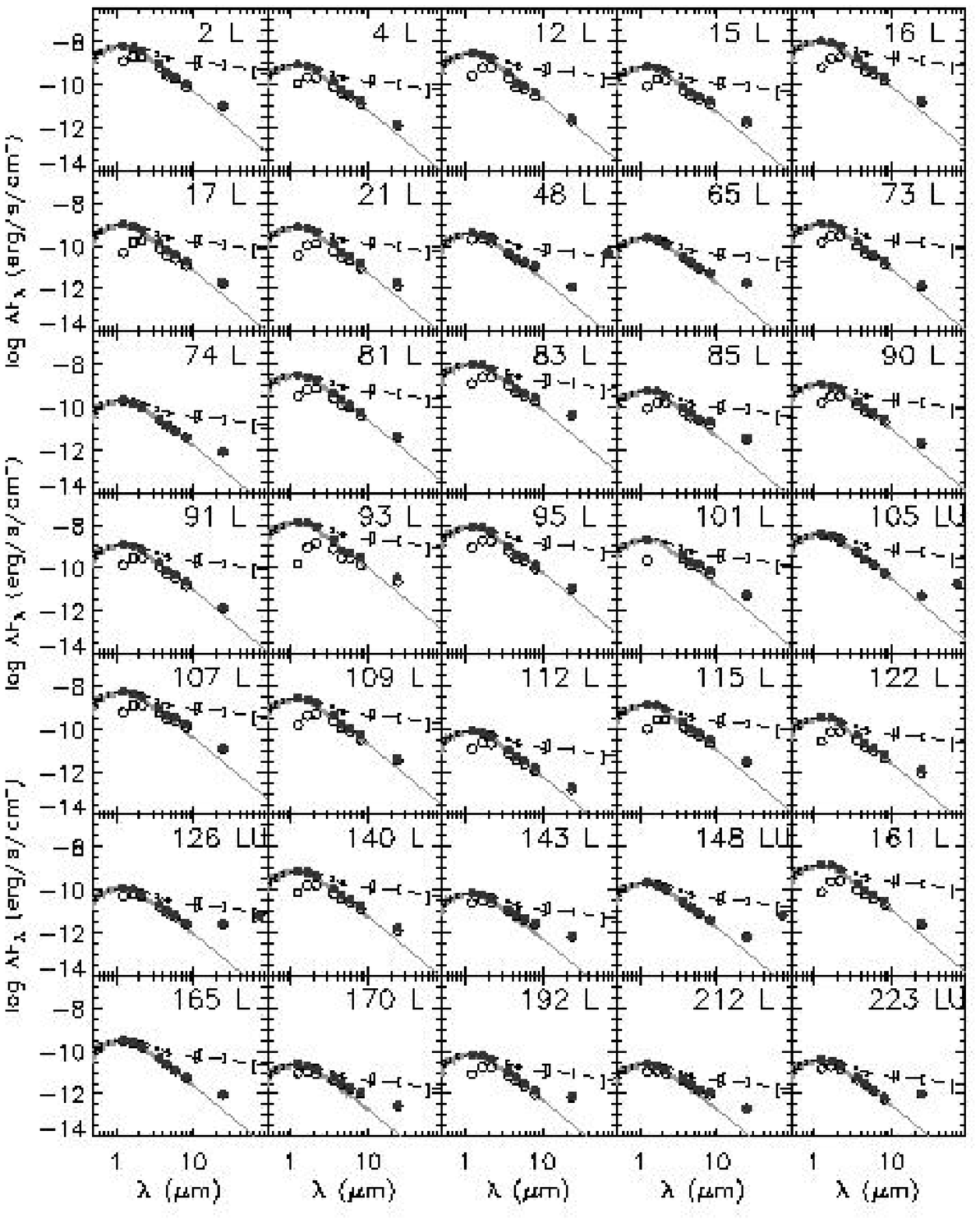}
\caption{\label{sedIII0}Spectral Energy Distributions of the Class III sources in the
sample.}
\end{figure}

\clearpage
\begin{figure}
\plotfiddle{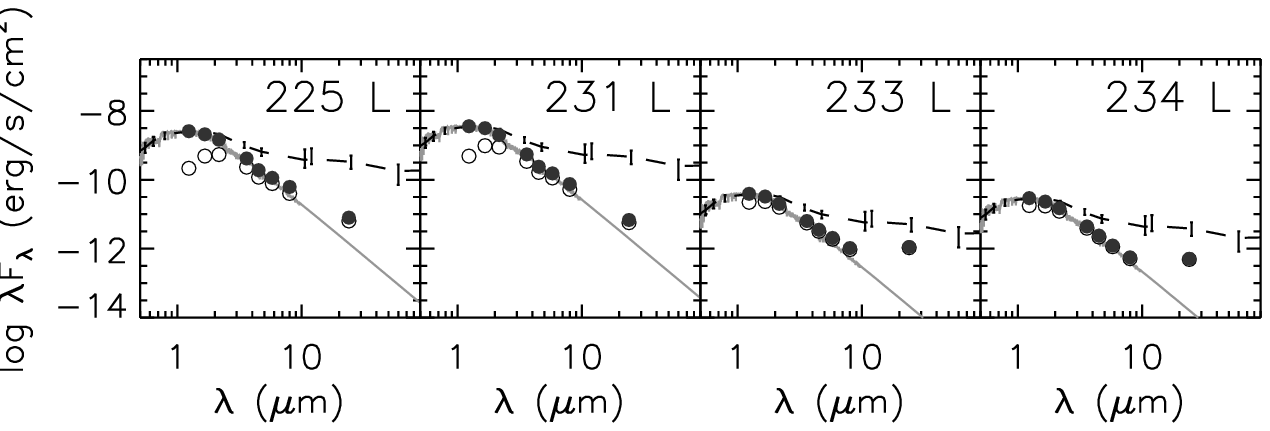}{4.0in}{0}{100}{100}{-180}{-400}
\caption{\label{sedIII1}Spectral Energy Distributions of the Class III sources in the
sample.}
\end{figure}
\clearpage

\begin{figure}
\includegraphics[angle=90,width=13cm]{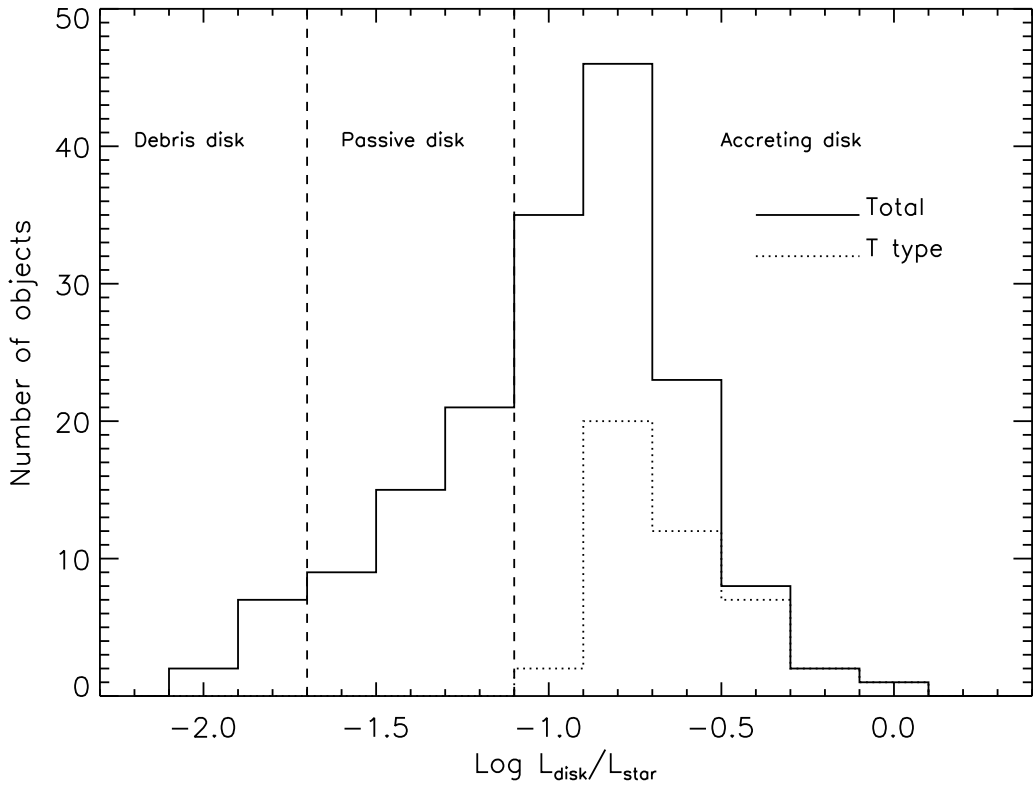}
\caption{\label{diskL}Distribution of disk to star luminosity ratios. The solid 
and dashed lines are the total sample and T Tauri-like sample of SEDs,
respectively. Also marked are the typical ranges of L$_{\rm
disk}$/L$_{\rm star}$ ratios for debris disks, passive irradiated 
disks and accretion disks. The figure indicates that objects of all
three evolutionary stages are found in Serpens, with a predominance
of young accreting T Tauri-type stars.}
\end{figure}

\begin{figure}
\plotfiddle{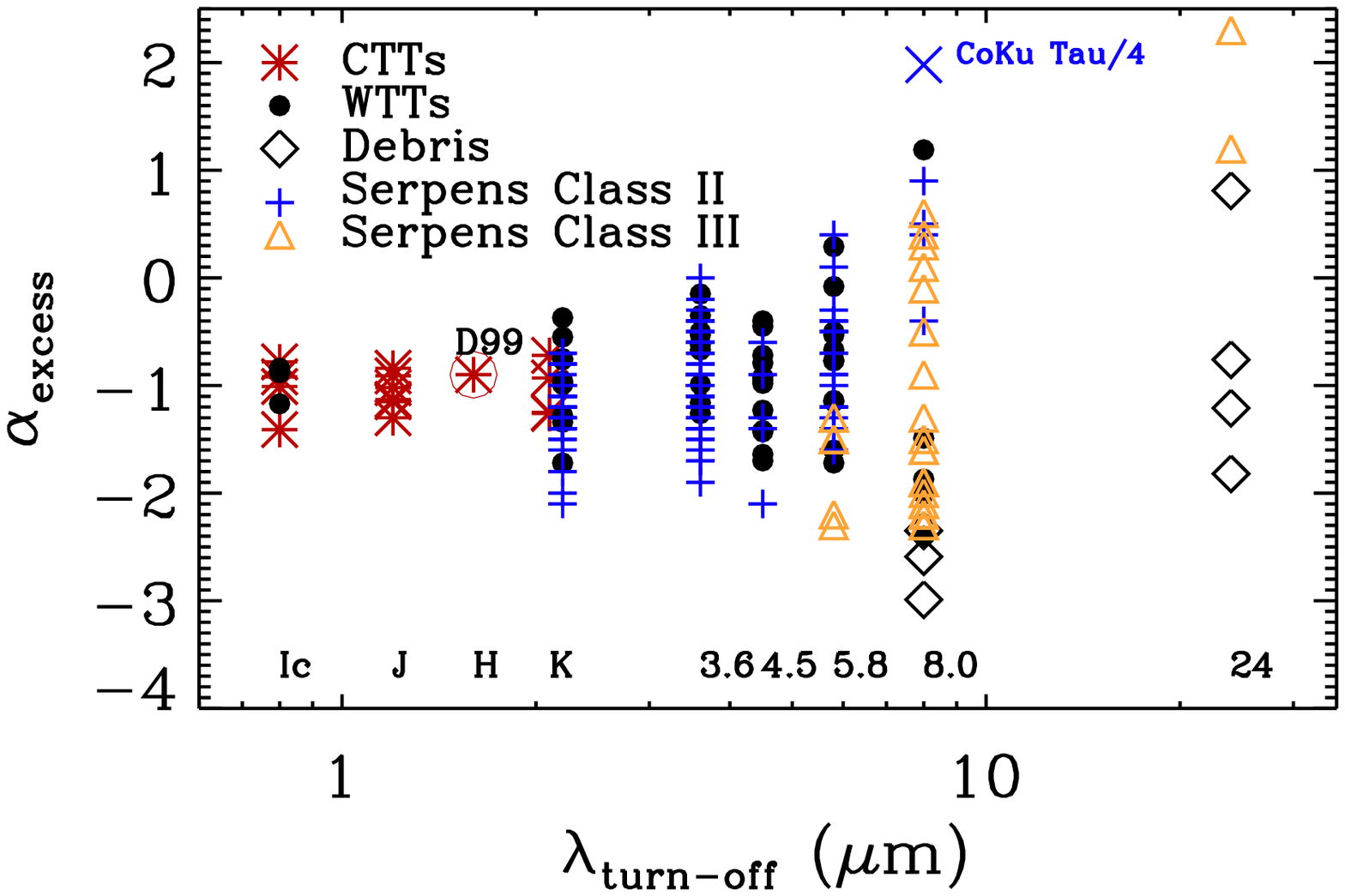}{5.0in}{0}{100}{100}{-280}{-300}
\caption{\label{alpha_turn}Distribution of excess slopes $\alpha_{\rm
   excess}$ with respect to the wavelength at which the infrared
   excess begins $\lambda_{\rm turn-off}$ for the sample of wTTs
   (solid dots), a sample of cTTs in Lupus from \citet{cieza06},
   the median SED of cTTs in Taurus from \citet{daless99}
   in asterisks (marked as D99), and a sample of debris disks from \citet{chen05}
 in diamonds. The arrows represent the limits for the
   class II objects in Serpens and the triangles the Class III
   sources. The diagram shows a much larger spread in inner disk
   morphologies in the more evolved objects than in the least evolved
   ones. The Serpens objects follow the previously observed trend.}
\end{figure}

\section{Selected Sources}\label{select}

\subsection{The Coldest Objects}\label{cold}

Table \ref{cold-table} lists the 15 YSO's that display the coldest energy distributions.  These
are the objects whose ratio of $F_{70}/F_{24} > 8$.  Six of these are in Cluster A (the core) and four
in Cluster B.
Two are in a small grouping about 0.2\degree\ northeast of Cluster B, and the remaining three appear as
isolated cold objects, though in areas with other moderately red objects. 
All of these except for two of the isolated cold objects, ID's 105 and 148, are associated
with dense clumps of millimeter emission as seen by \citet{enoch07}.
As discussed in the following section, two in Cluster A and one in Cluster B are associated with high velocity
outflows. 
As has been noted already by \citet{padgett04} and \citet{rebull06}, it is striking how many of these
coldest, most obscured YSO's are located in compact clusters together with objects that are substantially more
evolved in the nominal system of classes, e.g. Class II and III.

\subsection{Outflow Sources}\label{outflows}

A number of recent Spitzer studies have found that high-velocity shocked outflows from young stars are
visible in IRAC images, typically strongest at 4.5\micron\ \citep{green1}.  We have examined our images for such
outflows as well as for correspondance with the lists of published HH objects in the region \citep{ze99}.
Table \ref{outflow-table} summarizes the results of this effort.  Two strong, obvious
jet-like features are seen in Cluster A and one in Cluster B.  In addition, as shown in Table \ref{outflow-table},
we find small extended features at the positions of most of the known HH objects in Serpens that were within
our covered area.  The two ``jets'' in Cluster A appear to originate from two of the most deeply embedded
objects in this cluster that are associated with the sub-mm sources SMM1 and SMM5 \citep{ts98}.  Both these outflows
are aligned roughly in the NW-SE direction and are visible on both sides of the central 24\micron\ likely
exciting sources.  In Cluster B there is one obvious jet-like feature extending mostly south of 
YSO \# 75.  There is also probably faint emission visible 30\arcsec\
to the north of the embedded source as well.  
\citet{harv07} discuss the Cluster B jet and several nearby Spitzer sources in more detail.
Figures \ref{clustera-jet} and \ref{clusterb-jet} show 3-color
images of Clusters A and B with the color tables chosen to make these jets most visible.  
In addition to the optical HH objects in the Serpens Clouds, there are a number of high-velocity molecular
outflows that have been mapped in Cluster A \citep{dav99}.  These maps present a confusing picture of outflows, and
because of the relatively larger beamwidth of the mm observations and close packing of infrared sources in this
cluster, it is very difficult to associate the radio outflows unequivocally with particular Spitzer sources.

\begin{figure}
\plotfiddle{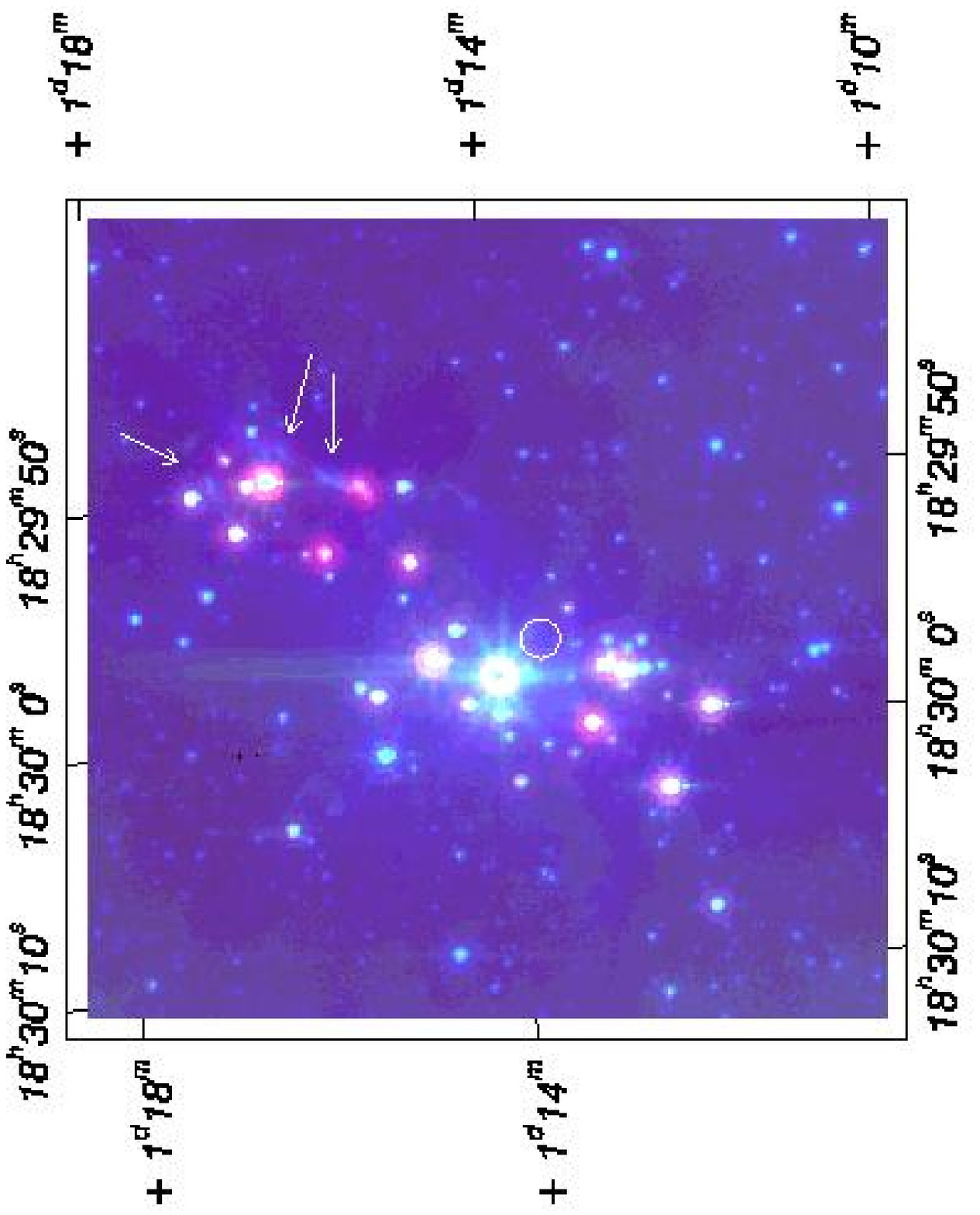}{7.0in}{-90}{90}{90}{-400}{500}
\figcaption{\label{clustera-jet}
Three-color image of the Cluster A area of Serpens.  The color coding is: blue/4.5\micron, green/8.0\micron, and
red/24\micron.  Some of the more obvious objects that are likely to be jets are marked by arrows.  The region of the
disappearing source 81 of \citet{ec92} is circled.}
\end{figure}

\begin{figure}
\plotfiddle{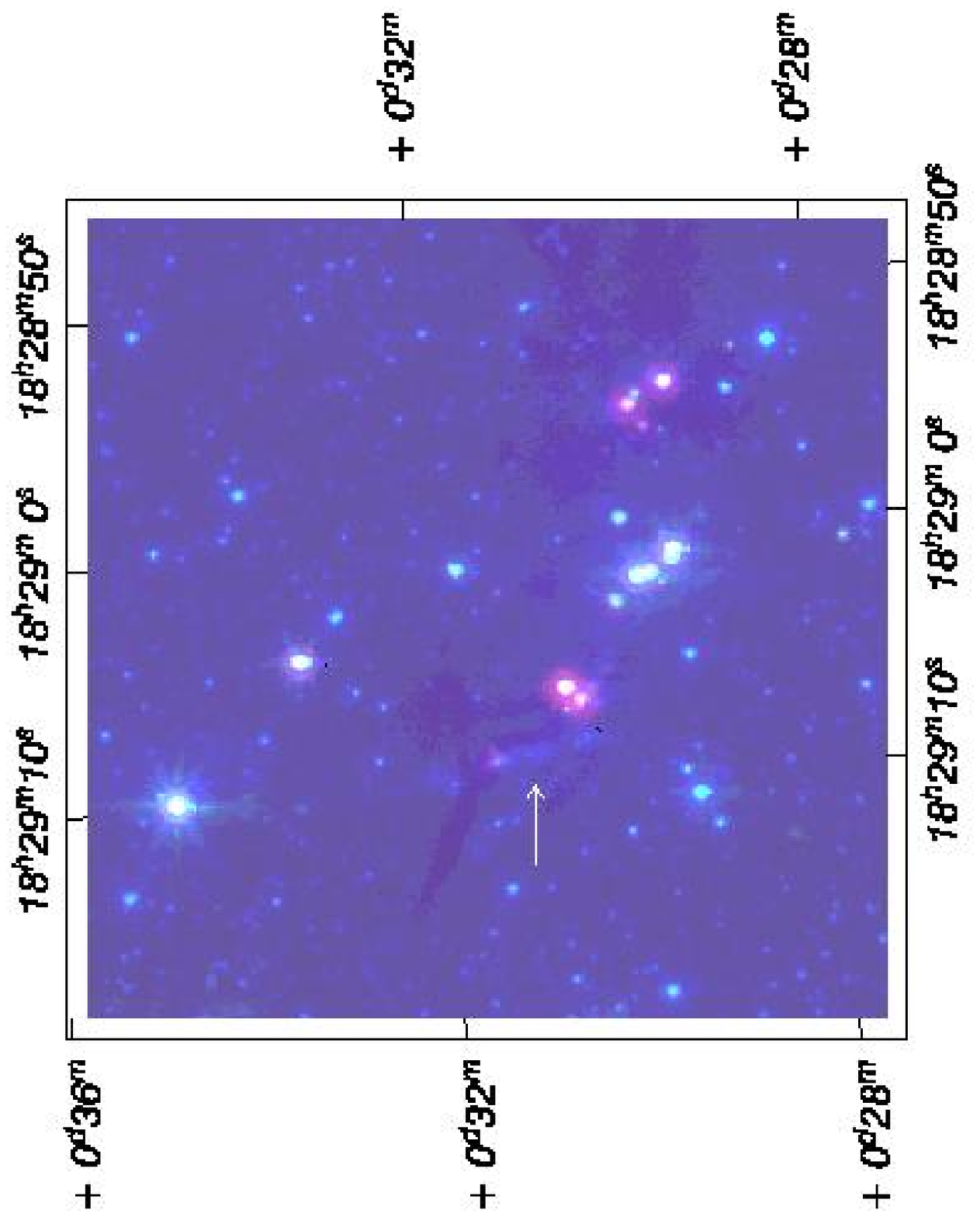}{7.0in}{-90}{90}{90}{-400}{500}
\figcaption{\label{clusterb-jet}
Three-color image of the Cluster B area of Serpens as for Figure \ref{clustera-jet}.}
\end{figure}

\subsection{Other Objects}

\citet{hce97} reported on the disappearance of a bright near-infrared source in
Serpens from the earlier study of EC92.  Table \ref{cross-corr} shows that we also see no obvious
source at the position of EC92-81, other than a low S/N single-band detection of a source moderately
distant from the nominal position.  Interestingly, though, this area marked in Figure \ref{clustera-jet} appears to contain several small
knots of emission that may represent shocked gas.  For example, there are two small knots visible in our
4.5 and 8.0\micron\ images that have no 3.6 or 24\micron\ counterparts (and therefore not classified as
YSO's).  These two knots at RA = 18 29 56.7, Dec = +01 13 19 (J2000), are only 12\arcsec\ south of the
position of EC92-81.  Therefore it is possible that the original source was a small clump of excited gas
that has moved or dissipated since the original study of EC92, and perhaps, these Spitzer knots are
related to that earlier source in some way.

\section{Overall Results on Star Formation}

The overall picture of star formation in Serpens is summarized in Table
\ref{tabyso}. As noted earlier, the surface density of young stars is
much higher in Clusters A and B than in the rest of the cloud, by factors of
10 to 20. On the other hand, the majority of YSOs (74\%) are {\bf not}
in the clusters, but in the rest of the cloud. If we assume a mean mass
for the stars of 0.5 \msun\ and assume that star formation has been proceeding
for 2 Myr, we can estimate the rate at which the clusters and the whole cloud
are converting mass into stars. These values are also given in 
Table \ref{tabyso}. The Serpens cloud that we have surveyed is turning
nearly 60 \msun\ into stars per Myr. The rates for the clusters are probably
underestimates because they are probably younger than 2 Myr, a typical
duration for the Class II SED.  Of course, a significant number of the
older objects outside the clusters may have, in fact, formed in these
or earlier clusters.  A 1 km s$^{-1}$ random motion of a YSO relative to
its birthplace results in 1 pc of movement in 10$^6$ yr, or nearly
1/4 degree at the distance of Serpens.

The distribution of YSOs over class supports the younger age for the
clusters than for star formation in general in Serpens. While the ratio
of the number of Class I and Flat spectrum sources to the number of
Class II and Class III sources is 0.37 for the whole cloud, similar to
other clouds surveyed by c2d (Evans et al., in prep.), 
the same ratio is 3.0 for Cluster A and 1.4
for Cluster B. These high ratios strongly suggest that Cluster A is too
young for most YSOs to have reached the Class II stage. In contrast, this
ratio is 0.14 for the rest of the cloud, outside the clusters, which is
strongly dominated by Class II and Class III objects.

\section{Summary}

We have identified a high-confidence set of 235 YSO's in Serpens by a set of criteria based on
comparison with data from the Spitzer SWIRE Legacy program.   This is a large enough number that
we can draw important statistical conclusions about various properties of this set.  If we
assume that the ``Class System'' of \citet{lada87} and \citet{greene94} represents an evolutionary
sequence, then the relative numbers of YSO's found in each of the first three classes  (I, flat, and II)
suggests that the Class II  phase lasts substantially longer than the combined 
total of the Class I and ``Flat'' phases, based on the overall cloud
statistics.  The clusters, however, show more YSOs in the Class I and Flat
phases than in Class II, indicating that they are very young.
The majority of YSOs (mostly Class II) are outside the clusters and
probably represent a somewhat older epoch of star formation compared to the
intense star formation now going on in the clusters. The surface density of
YSOs in the clusters exceeds that of the rest of the cloud by factors of
10 to 20 (Table \ref{tabyso}).

The luminosity function for the Serpens YSO's is populated down to 
luminosities of
$3 \times 10^{-3}$ \lsun. It may extend even lower, but our ability to distinguish low luminosity
YSO's is severely restricted by the large population of background galaxies at these flux levels.
The lower luminosity YSO's, $L < 2 \times 10^{-2}$ \lsun, exhibit a similar spatial distribution
and SED slope distribution to those of their higher luminosity counterparts.  This is consistent
with the conclusion that they have formed in similar ways.  

\section{Acknowledgments}

Support for this work, part of the Spitzer Legacy Science
Program, was provided by NASA through contracts 1224608, 1230782, and
1230779 issued by the Jet Propulsion Laboratory, California Institute
of Technology, under NASA contract 1407. 
Astrochemistry in Leiden is supported by a NWO Spinoza grant
and a NOVA grant.  B.M. thanks the Fundaci\'on Ram\'on Areces for financial support.
This publication makes use of data products from the
Two Micron All Sky Survey, which is a joint project of the University
of Massachusetts and the Infrared Processing and Analysis
Center/California Institute of Technology, funded by NASA and the
National Science Foundation.  We also acknowledge extensive use of the SIMBAD
data base.

\clearpage
\appendix \label{class_math}
\section{YSO Selection Process}

The procedure we use to select YSO's and de-select extra-galactic background sources is based on
the color-magnitude diagrams shown in Figure \ref{cm_all}.   We construct ``probability'' functions
for each of the three color-magnitude diagrams based on where a source falls relative to the
black dashed lines in each diagram.  These three ``probabilities'' are multiplied and then additional
adjustments to the probability are made based on several additional properties of the source fluxes
and whether or not they were found to be larger than point-like in the source extraction process.

In the [4.5] vs. [4.5]-[8.0] color magnitude diagram, the probability function is:
$$P_{i2i4} = 0.7 \times \{1 - exp(-[1.2 + 0.5(M_{4.5}-D)]^3)\}$$
where:
\begin{eqnarray*}
 D &=& 13.05 \mbox{ for } M_{4.5}-M_{8.0} > 1.4 \nonumber \\
  &=& 14.5 \mbox{ for } 0.5 \ge M_{4.5}-M_{8.0} < 1.4 \mbox{ and object was found to be pointlike in all IRAC bands} \nonumber \\
  &=& 12.5 \mbox{ for } 0.5 \ge M_{4.5}-M_{8.0} < 1.4 \mbox{ and object was found to be extended at 3.6 or 4.5\micron} \nonumber\\
  &=& 14.5 \mbox{ for } M_{4.5}-M_{8.0} < 0.5  \nonumber \\
\end{eqnarray*}

In the [24] vs. [8.0]-[24] color magnitude diagram, the probability function is:
$$P_{i4m1} =  exp(-\{[(M_8 - M_{24}-3.5)/1.7]^2 +[(M_{24}-9)/2)^2\}$$
and $P_{i4-m1}$ is set to a minimum of 0.1.

In the [24] vs. [4.5]-[8.0] color magnitude diagram, the probability function is:
$$P_{i24m1} = 0.7 \times \{1 - exp(-[M_{4.5}-M_8 +0.8 -(10 - M_{24})/1.8 \}$$
and $P_{i24m1}$ is set to a minimum of 0.

The combined ``probability'' is then:
$$P_{tot} =  P_{i2i4} \times P_{i4m1} \times P_{i24m1}$$
Additionally, the following factors influence modifications to $P_{tot}$:
\begin{eqnarray*}
P_{tot} &/& (K - [4.5]) \\
P_{tot} &\times& 2 \mbox{ for sources that are extended at 3.6 or 4.5\micron} \nonumber \\
P_{tot} &=& 0.1 \mbox{ where [24] $>$ 10 (i.e. assumed GALc)} \nonumber \\
P_{tot} &\times& 0.1 \mbox{ where $F_{70} >$ 400 mJy and source detected at 5.8,8, and 24\micron} \nonumber \\
\end{eqnarray*}

Based on the distribution of $P_{tot}$ shown in Figure \ref{probcut}, we chose the dividing line between
YSO's and extra-galactic sources to be $log_{10}(P_{tot}) = -1.47$.


\begin{deluxetable}{ll}
\tablecolumns{2}
\tablecaption{YSO vs. X-Gal Discrimination Criteria\label{prob_table}}
\tablewidth{0pt}
\tablehead{
\colhead{Criterion}                &
\colhead{Contaminant Probability}               \\
}
\startdata
 K - [4.5] & Prob$ / (K - [4.5])$ \cr
 [4.5] (for [4.5]-[8.0] $<$ 0.5) & Smooth increase for [4.5] $>$ 14.5 \cr
 [4.5] (for 0.5 $<$ [4.5]-[8.0] $<$ 1.4, Extended) & Smooth increase for [4.5] $>$ 12.5 \cr
 [4.5] (for 0.5 $<$ [4.5]-[8.0] $<$ 1.4, Pointlike) & Smooth increase for [4.5] $>$ 14.5 \cr
 [4.5] (for 1.4 $<$ [4.5]-[8.0]) & Smooth increase for [4.5] $>$ 13.0 \cr
 $([8.0]-[24]-3.5)^2/1.7 + ([24]-9.0)^2/2$ & Smooth increase for values $<$ 1 \cr
 $[4.5]-[8.0]-0.55(10.-[24])$ & Smooth increase for values $>$ 1 \cr
 [24] (for all colors) & 100\% probability for [24] $>$ 10.0 \cr
Extended at 3.6 or 4.5\micron\ & Twice as likely to be X-gal \cr
$F_{70}$ & X-gal prob $\times$ 0.1 for $F_{70} >$ 400 mJy \cr
\enddata
\end{deluxetable}

\begin{deluxetable}{lcccccccc}
\tabletypesize{\small}
\rotate
\tablecolumns{8}
\tablecaption{YSO's in Serpens\label{yso-table}}
\tablewidth{0pc} 
\tablehead{
\colhead{ID}                &
\colhead{Name/Position} &
\colhead{Prev. Name\tablenotemark{a}} &
\colhead{3.6 \micron\ }  &
\colhead{4.5 \micron\ }  &
\colhead{5.8 \micron\ }  &
\colhead{8.0 \micron\ }  &
\colhead{24.0 \micron\ }  &
\colhead{70.0 \micron\ }               \\
\colhead{}                &
\colhead{SSTc2dJ...} &
\colhead{}                &
\colhead{(mJy)}  &
\colhead{(mJy)}  &
\colhead{(mJy)}  &
\colhead{(mJy)}  &
\colhead{(mJy)}  &
\colhead{(mJy)}                  
}

\startdata 

  1 & 18275381$-$0002333 &                   & 25.5$\pm$ 1.2 & 20.2$\pm$ 1.0 & 15.6$\pm$ 0.7 & 13.9$\pm$ 0.7 & 22.7$\pm$ 2.1 & \nodata\\
  2\tablenotemark{b} & 18280503$+$0006593 &                   &  783$\pm$  55 &  426$\pm$  29 &  358$\pm$  18 &  218$\pm$  14 & 76.4$\pm$ 7.1 & \nodata\\
  3 & 18280845$-$0001064 &                   &  140$\pm$   7 &  108$\pm$   8 &  129$\pm$   6 &  122$\pm$   8 &  161$\pm$  14 &  314$\pm$  36\\
  4\tablenotemark{b} & 18281100$-$0001393 &                   &  104$\pm$   5 & 63.4$\pm$ 3.2 & 54.6$\pm$ 2.6 & 35.4$\pm$ 1.9 & 9.51$\pm$0.90 & \nodata\\
  5 & 18281350$-$0002491 & CDF88-2           &  105$\pm$   5 & 88.1$\pm$ 4.7 & 77.5$\pm$ 3.7 & 94.3$\pm$ 5.0 &  254$\pm$  23 &  382$\pm$  43\\
  6 & 18281501$-$0002588 &                   & 33.3$\pm$ 1.6 & 45.6$\pm$ 2.3 & 61.5$\pm$ 3.1 & 96.3$\pm$ 5.8 &  225$\pm$  20 &  320$\pm$  41\\
  7 & 18281519$-$0001405 &                   & 7.18$\pm$0.35 & 6.18$\pm$0.30 & 5.15$\pm$0.25 & 5.01$\pm$0.26 & 8.00$\pm$0.77 & \nodata\\
  8 & 18281525$-$0002432 & CoKu Ser-G1       &  174$\pm$   9 &  177$\pm$   9 &  180$\pm$  11 &  191$\pm$  11 & 1200$\pm$   25 & 1670$\pm$  169\\
\enddata

\tablenotetext{a}{Source names from SIMBAD, including numbers from the following catalogs: EC, \citet{ec92}; D, \citet{djup06}; K, \citet{kaas04}.}
\tablenotetext{b}{May be AGB star, based on $A_v$ derived from optical spectrum.}
\end{deluxetable}

\begin{deluxetable}{lcccccccc}
\tabletypesize{\small}
\rotate
\tablecolumns{8}
\tablecaption{YSO Candidates in Serpens Without 4-Band IRAC Observations\label{mipsyso_table}}
\tablewidth{0pc} 
\tablehead{
\colhead{ID}                &
\colhead{Name/Position} &
\colhead{Prev. Name} &
\colhead{3.6 \micron\ }  &
\colhead{4.5 \micron\ }  &
\colhead{5.8 \micron\ }  &
\colhead{8.0 \micron\ }  &
\colhead{24.0 \micron\ }  &
\colhead{70.0 \micron\ }               \\
\colhead{}                &
\colhead{SSTc2dJ...} &
\colhead{}                &
\colhead{(mJy)}  &
\colhead{(mJy)}  &
\colhead{(mJy)}  &
\colhead{(mJy)}  &
\colhead{(mJy)}  &
\colhead{(mJy)}                  
}

\startdata 

236 & 18273421$+$0040260 &                   & \nodata & \nodata & \nodata & \nodata & 17.8$\pm$ 1.7 & \nodata\\
237 & 18274024$+$0035221 &                   & \nodata & \nodata & \nodata & \nodata & 1.36$\pm$0.25 & \nodata\\
238 & 18274264$+$0020389 &                   & \nodata & \nodata & \nodata & \nodata & 42.0$\pm$ 3.9 & \nodata\\
239 & 18274399$+$0025135 & IRAC 18251+0023   & \nodata & \nodata & \nodata & \nodata & 5430$\pm$  509 &  655$\pm$  68\\
240 & 18274484$+$0052436 &                   & \nodata & \nodata & \nodata & \nodata & 1.49$\pm$0.27 & \nodata\\
241 & 18274506$+$0006110 & IRAS 18251+0004   & 1730$\pm$   93 & \nodata &  965$\pm$  47 & \nodata &  173$\pm$  16 & \nodata\\
242 & 18275926$+$0056012 &                   & \nodata & \nodata & \nodata & \nodata & 40.7$\pm$ 3.9 & \nodata\\
243 & 18280449$+$0049498 &                   & \nodata & \nodata & \nodata & \nodata & 15.2$\pm$ 1.4 & \nodata\\
\enddata
\end{deluxetable}

\begin{deluxetable}{lccccccc}
\rotate
\tablecolumns{8}
\tablecaption{Matching Spitzer Sources/Fluxes For Previously Identified IR Sources In Serpens\label{cross-corr}}
\tablewidth{0pt} 
\tablehead{
\colhead{Source\tablenotemark{a}} &
\colhead{Spitzer} &
\colhead{3.6\micron}  &
\colhead{4.5\micron}  &
\colhead{5.8\micron}  &
\colhead{8.0\micron}  &
\colhead{24\micron} &
\colhead{70\micron} \\

\colhead{} &
\colhead{SSTc2dJ...} &
\colhead{(mJy)}  &
\colhead{(mJy)}  &
\colhead{(mJy)}  &
\colhead{(mJy)}  &
\colhead{(mJy)}  &
\colhead{(mJy)} 
}

\startdata 

  D002 & 18282432$+$0034545 & 62.4$\pm$ 4.2 & 46.1$\pm$ 3.0 & 41.1$\pm$ 2.5 & 32.5$\pm$ 2.0 & 13.9$\pm$ 1.3 & \nodata\\
  D001 & 18282442$+$0026509 &  105$\pm$   9 & 95.6$\pm$ 7.8 & 97.9$\pm$ 6.5 & 65.0$\pm$ 4.2 & 13.9$\pm$ 1.3 & \nodata\\
  D003 & 18282649$+$0031421 & 38.1$\pm$ 2.7 & 25.8$\pm$ 1.7 & 20.1$\pm$ 1.2 & 12.2$\pm$ 0.8 & 1.26$\pm$0.28 & \nodata\\
  D004 & 18282783$+$0028390 & 6.72$\pm$0.45 & 4.47$\pm$0.29 & 3.44$\pm$0.21 & 2.27$\pm$0.14 & $<$  1.53 & \nodata\\
  D005 & 18282792$+$0029288 & 5.12$\pm$0.32 & 3.56$\pm$0.25 & 2.76$\pm$0.16 & 1.79$\pm$0.12 & $<$  1.45 & \nodata\\
  D006 & 18282864$+$0024120 & 5.68$\pm$0.38 & 3.71$\pm$0.25 & 3.27$\pm$0.20 & 1.92$\pm$0.12 & $<$  1.44 & \nodata\\
  D007 & 18282905$+$0027561 & 9.80$\pm$0.57 & 8.51$\pm$0.47 & 7.03$\pm$0.38 & 5.46$\pm$0.30 & 43.4$\pm$ 4.0 & \nodata\\
  D008 & 18282912$+$0034193 &  193$\pm$  13 & 93.2$\pm$ 6.4 & 86.6$\pm$ 4.5 & 49.3$\pm$ 2.9 & 5.99$\pm$0.58 & \nodata\\
\enddata

\tablenotetext{a}{Source numbers from the following catalogs: EC, \citet{ec92}; D, \citet{djup06}; P, \citet{preib03}, K, \citet{kaas04}.}
\end{deluxetable}

\begin{deluxetable}{ccc}
\tablecolumns{3}
\tablecaption{Average/Std Dev of Spectral Slope $\alpha$ Versus Luminosity of YSO's\label{alpha_lum_table}}
\tablewidth{0pc} 
\tablehead{
\colhead{Luminosity Range}                &
\colhead{Average $\alpha$}  &
\colhead{$\sigma_\alpha$}
}
\startdata 
$< 0.02 $ \lsun & -0.73 & 1.47 \\
$0.02 -- 1.0 $ \lsun & -0.70 & 1.04 \\
$> 1.0 $ \lsun & -0.60 & .74 \\
\enddata
\end{deluxetable}

\begin{deluxetable}{lccccccc}
\tablecolumns{8}
\tablecaption{Numbers, Densities, Star Formation Rates \label{tabyso}}
\tablewidth{0pt}
\tablehead{
\colhead{Region}  & \colhead{Area} &  \colhead{N(YSOs)} & \colhead{N/$\Omega$}  & \colhead{N/Area} & \colhead{N/Vol} & \colhead{SFR}  & \colhead{SFR/Area}  \\
 &  & \colhead{(pc$^2$)} & \colhead{(deg$^{-2}$)}  & \colhead{(pc$^{-2}$)} & \colhead{(pc$^{-3}$)} & \colhead{(\msunmyr)} & \colhead{(\msunmyr pc$^{-2}$)}    \\
}
\startdata
Cluster A & 0.20	& 44	& 4580	& 222 & 500 & 11   & 56   \cr
Cluster B & 0.14	& 17	& 2450	& 119 & 315 & 4.3  & 30   \cr
Rest of Cloud & 17.2    & 174	& 209	& 10.1 & 2.5 & 44   & 2.5   \cr
All of Cloud  & 17.5	& 235 	& 276	& 13.4 & 3.2 & 59   & 3.4   \cr
\enddata
\end{deluxetable}

\begin{deluxetable}{lllllll}
\tabletypesize{\small}
\tablecolumns{7}
\tablecaption{Disk Modeling Results in Class II Sources\label{resultsII}}
\tablewidth{0pc}
\tablehead{
\colhead{Star}                &
\colhead{SED Type}  &
\colhead{A$_V$}     &
\colhead{$L_{star}$} &
\colhead{$L_{disk}/L_{star}$} &
\colhead{$\lambda_{\rm turn-off}$} &
\colhead{$\alpha_{\rm excess}$}
}

\startdata
   1 &   L &   2.9 &   0.33 &  0.088 &  5.8 &  -0.7 \\
   3 &   T &   4.6 &   1.06 &  0.330 &  2.2 &  -0.8 \\
   5 &   T &   4.9 &   1.64 &  0.165 &  3.6 &  -0.5 \\
   7 &   L &   3.8 &   0.09 &  0.110 &  3.6 &  -1.0 \\
   9 &   L &   6.6 &   0.91 &  0.141 &  3.6 &  -0.8 \\
  10 &   L &   9.9 &   0.05 &  0.190 &  3.6 &  -1.3 \\
  11 &   L &   8.4 &   0.07 &  0.186 &  2.2 &  -1.2 \\
  13 &   L &   2.8 &   0.19 &  0.084 &  5.8 &  -1.0 \\
\enddata
\end{deluxetable}

\begin{deluxetable}{lllllll}
\tabletypesize{\small}
\tablecolumns{7}
\tablecaption{Disk Modeling Results in Class III Sources\label{resultsIII}}
\tablewidth{0pc}
\tablehead{
\colhead{Star}                &
\colhead{SED Type}  &
\colhead{A$_V$}     &
\colhead{$L_{star}$} &
\colhead{$L_{disk}/L_{star}$} &
\colhead{$\lambda_{\rm turn-off}$} &
\colhead{$\alpha_{\rm excess}$}
}

\startdata

   2 &   L &   6.3 &  19.19 &  0.013 & 24.0 & -99.0 \\
   4 &   L &   7.6 &   2.50 &  0.028 & 24.0 & -99.0 \\
  12 &   L &   9.1 &   8.76 &  0.015 & 24.0 & -99.0 \\
  15 &   L &   7.7 &   2.02 &  0.025 &  8.0 &  -1.9 \\
  16 &   L &  10.8 &  31.18 &  0.041 &  8.0 &  -2.3 \\
  17 &   L &  11.1 &   2.69 &  0.036 &  8.0 &  -2.1 \\
  21 &   L &  11.7 &   2.16 &  0.038 &  8.0 &  -2.0 \\
  48 &   L &   2.5 &   1.13 &  0.054 &  8.0 &   0.6 \\
\enddata
\end{deluxetable}

\begin{deluxetable}{lccccccc}
\tabletypesize{\small}
\rotate
\tablecolumns{7}
\tablecaption{The Coldest YSO's ($F_{70}/F_{24} > 8$)\label{cold-table}}
\tablewidth{0pc} 
\tablehead{
\colhead{ID}                &
\colhead{Name/Position} &
\colhead{3.6 \micron\ }  &
\colhead{4.5 \micron\ }  &
\colhead{5.8 \micron\ }  &
\colhead{8.0 \micron\ }  &
\colhead{24.0 \micron\ }  &
\colhead{70.0 \micron\ }               \\
\colhead{Tbl \ref{yso-table}}                &
\colhead{SSTc2dJ...} &
\colhead{(mJy)}  &
\colhead{(mJy)}  &
\colhead{(mJy)}  &
\colhead{(mJy)}  &
\colhead{(mJy)}  &
\colhead{(mJy)}                  
}

\startdata 

 40 & 18285404$+$0029299 & 5.81$\pm$0.50 & 27.6$\pm$ 2.3 & 44.8$\pm$ 2.6 & 56.4$\pm$ 3.2 &  918$\pm$  85 & 11100$\pm$  1040\\
 42 & 18285486$+$0029525 & 1.94$\pm$0.12 & 10.6$\pm$ 0.6 & 20.4$\pm$ 1.1 & 30.2$\pm$ 1.6 &  765$\pm$  70 & 7250$\pm$  675\\
 48 & 18285808$+$0017244 & 52.5$\pm$ 3.0 & 36.7$\pm$ 2.5 & 31.2$\pm$ 1.7 & 28.5$\pm$ 1.7 & 9.74$\pm$0.92 & 1040$\pm$  101\\
 60 & 18290211$+$0031206 & 1.19$\pm$0.07 & 1.62$\pm$0.09 & 1.58$\pm$0.10 & 1.13$\pm$0.07 & 22.1$\pm$ 2.0 &  276$\pm$  29\\
 68 & 18290675$+$0030343 & 3.27$\pm$0.21 & 11.7$\pm$ 0.7 & 14.9$\pm$ 0.8 & 20.7$\pm$ 1.2 & 1000$\pm$  105 & 11400$\pm$  1180\\
105 & 18293254$-$0013233 &  654$\pm$  39 &  370$\pm$  19 &  276$\pm$  13 &  156$\pm$   8 & 39.7$\pm$ 3.7 &  429$\pm$  47\\
135 & 18294810$+$0116449 & 1.96$\pm$0.10 & 6.98$\pm$0.42 & 12.1$\pm$ 0.6 & 16.7$\pm$ 0.8 &  219$\pm$  21 & 14900$\pm$  1420\\
141 & 18294963$+$0115219 & 0.85$\pm$0.08 & 2.64$\pm$0.27 & 2.32$\pm$0.28 & 3.54$\pm$0.31 & 1180$\pm$  117 & 82800$\pm$  7810\\
\enddata

\end{deluxetable}

\begin{deluxetable}{lcccc}
\tablecolumns{5}
\tablecaption{Probable High Velocity Outflows in Serpens\label{outflow-table}}
\tablewidth{0pc} 
\tablehead{
\colhead{YSO ID}                &
\colhead{RA (J2000)}  &
\colhead{Dec (J2000)}     &
\colhead{Matching Name} &
\colhead{Comments}
}
\startdata 
     &  18 29 18.8 & +01 14 15 & HH 106A,B,E &  \\
     &  18 29 47.7 & +01 25 52 & HH 107A &  \\
     &  18 30 22.7 & +01 16 18 & HH 455A &  \\
     &  18 30 22.6 & +01 16 05 & HH 455B,D &  \\
     &  18 30 02.6 & +01 14 45 & HH 459A,B &  \\
     &  18 29 38.5 & +01 18 26 & HH 460A,B,C,D &  \\
     &  18 29 56.6 & +01 15 37 & HH 478A,B &  \\
141  &  18 29 49.6 & +01 15 21 & SMM 1 & Jets to NW and SE \\
146  &  18 29 51.2 & +01 16 41 & SMM 5 & Jets to NW and SE \\
75   &  18 29 09.0 & +00 31 30 &  & Jets to N and S \\
\enddata
\end{deluxetable}


\clearpage

\end{document}